\documentclass{jaa}
\usepackage{natbib}
\usepackage{graphicx}

\usepackage{microtype}
\begin{document}

\title{Coronal spectroscopy\,--\,Past, Present and Future }

\author{Jagdev Singh, R. Ramesh, V. Muthu Priyal}

\affilOne{Indian Institute of Astrophysics, Bangalore 560 034, Karnataka, India.\\}

\twocolumn[{

\maketitle

\corres{jsingh@iiap.res.in}

\begin{abstract}
		
	After the beginning of photography of solar corona during the eclipses in the year 1860, efforts began to carry out spectroscopic observations of solar corona. During the total solar eclipse of 1868 August 18 at Guntur, Madras state, the observation of emission line 5875{\AA} by Janssen lead to the discovery of Helium. Next year, during the eclipse of 1869 August 7, Charles Young and William Harkness discovered green emission line around 5310{\AA} (later measured as 5303{\AA}). It was interpreted as due to Coronium, an unknown element. The identification of this green line in 1939 due to [Fe xiv] ion showed the corona to be hot at ${\sim}10^{6}$\,K. High resolution spectroscopic observations indicated that width of emission lines to be more than expected. The excess line width was interpreted in terms of existence of non-thermal velocity. Recent spectroscopic observations have indicated a complex nature of variation of emission line parameters with height above the corona. The spectroscopy of solar corona is also performed in other wavelengths, such as UV, EUV, X-rays, visible, infrared, and radio part of the spectrum. Here, we discuss the past spectroscopic observations of solar corona, present observations with various instruments, such as Visible Emission Line Coronagraph onboard Aditya-L1, and need for the spectroscopic observations in future.

	\end{abstract}
	
}]

\pgrange{1--}
\setcounter{page}{1}
\lp{19}

\section{Introduction}

Spectroscopy of the solar corona is important as it provides information about the temperature, velocity and density of the plasma whereas images of solar corona provide information about physical structure and their brightness only. Spectroscopy also helps to identify the elements and their abundances in the plasma. Thus, spectroscopy is essential to understand the solar coronal physical and dynamical characteristics in detail. In 1860  American astronomer Charles A. Young used wet photographic plates to take the images of solar corona during an eclipse for the first time. The technique to take the photographs developed fast and was applied to do spectroscopy of solar corona during eclipses. Spectroscopy began when Newton, in the beginning of $18^{\mathrm{th}}$ century, described how the use of a prism of solid glass splits an incoming ray of white light into its colors, the blue colour being refracted more than the red colour.  He used the word ``Spectrum'' to describe the resulting colors from blue to red when projected on a wall. French astronomer Pierre Janssen made spectroscopic observations during the eclipse of 1868 August 18 from Guntur in Madras State. He noticed a bright yellow line at a wavelength of 5875{\AA} in the spectra of the solar prominences. The line was not due to sodium as had been assumed previously. The same result was found independently by British astronomer  Norman Lockyer from the observations during the same eclipse. These observations lead to the discovery of Helium. Janssen argued that the line is very bright and should be possible to observe without the total solar eclipse. Next day he pointed the telescope at the same location and was able to observe the same emission line even without the need for an eclipse. Thereafter, in the beginning of $20^{\mathrm{th}}$ century, observatories such as Mount Wilson, Greenwich, and Kodaikanal in India started regular study of prominences. The spectroscopy of the solar corona is a vast subject. It is performed at wavelengths from X-rays, EUV, UV, visible, near and far Infra-red, mm wave to radio waves. A large number of instruments onboard various satellites have been used to perform spectroscopy in the X-rays, EUV and UV wavelength regions. We shall describe in brief few of the instruments.
The Japanese satellite Yohkoh is well known for its soft X-ray and hard X-ray telescopes (SXT and HXT) to take images of Sun, thereby of solar corona as the background emission from the solar disk is significantly reduced, allowing the much fainter corona to be observed. It also had two spectrometers; (1) Bragg crystal spectrometer (BCS) to perform spectroscopic diagnosis of ultra high-temperature plasma observing in the range 1\,-\,5{\AA} by integrating the radiation over the solar disc and; (2) Wide-band spectrometer (WBS) to conduct spectroscopic observations of the 
wide energy region from soft X-rays to gamma rays, in a wide energy band from 3 keV to 100 MeV. The Coronal Diagnostic Spectrometer (CDS) onboard Solar and Heliospheric Observatory (SoHO) observed in the EUV wavelength range 150\,-\,800{\AA} where there are many emission lines of highly stripped ions of characteristic temperatures 
$1\,{\times}\,10^{4}$\,K\,-\,2\,${\times}\,10^{6}$\,K. The analysis of data provided density and temperature diagnostics of the transition region and low corona \citep{harrison1995}.    
Solar Ultraviolet Measurements of Emitted Radiation (SUMER) onboard SoHO has the facility to perform the coronal spectroscopy in the wavelength range of 500\,-\,1600{\AA}. This facilitate to measure plasma flows, temperature, and density in the corona. The EUV Imaging Spectrometer (EIS) onboard the Hinode mission has the facility to take the spectra of the Sun in wavelength band of 170\,-\,210{\AA} and 250\,-\,290{\AA} \citep{watanabe2007}. In addition, the spectroscopic observations with X-rays spectrometer SOLFLEX (1979\,-\,1981), Solar Maximum Mission (1980, 1984\,-\,1989), Yohkoh (1991\,-\,2001), RHESSI (2002\,-\,2018), Indian mission SOXS on GSAT-2 (2004\,-\,2011), SoLEXS and HEL1OS onboard Aditya-L1 \citep{parate2025} and many others have contributed to understand the physics of high energy events.  
In an ionized medium like the corona, radiation shifts to lower frequencies with increasing height. Radio observations at frequencies $<$\,150 MHz help to map the outer corona in the heliocentric distance range $\approx$1.2\,-\,3$R_{\odot}$. The spectral observations target non-thermal bursts, serving as vital diagnostics for coronal electron density, magnetic fields, space weather, and particle acceleration.  Some of the major low frequency solar radio spectrographs currently in operation are e-CALLISTO \citep{benz2009}, Gauribidanur Low-frequency Solar Spectrograph \citep{ramesh2001,ramesh2011}, Yamagawa solar radio spectrograph \citep{iwai2017}, Radio Solar Telescope Network spectrographs \citep{giersch2017}. Some of the imaging instruments for dedicated observations of the solar corona at low frequencies are Gauribidanur Radioheliograph \citep{ramesh1998,hariharan2014}, Nancay radioheliograph \citep{kerdraon1997}, and Mingantu Spectral Radioheliograph for Solar and Space Weather Studies \citep{yan2021}. Interestingly, solar eclipses provide an excellent opportunity to carry out high angular resolution (${\approx}$15”) radio observations at low frequencies, similar to the lunar occultation technique \citep{Ramesh1999,Ryan2021}.
There are many instruments to perform the spectroscopy in various wavelength bands on ground and in space. Here, we shall discuss the spectroscopy performed in visible wavelength region.  We have considered the spectroscopy of solar corona done up to $20^{\mathrm{th}}$ century as the past period and in the $21^{\mathrm{st}}$ century as the present period.

\section{Coronal spectroscopy in the past\,--\,up to $20^{\mathrm{th}}$ century}

\begin{figure*}[!ht]
	\centering
	\includegraphics[width=0.95\linewidth]{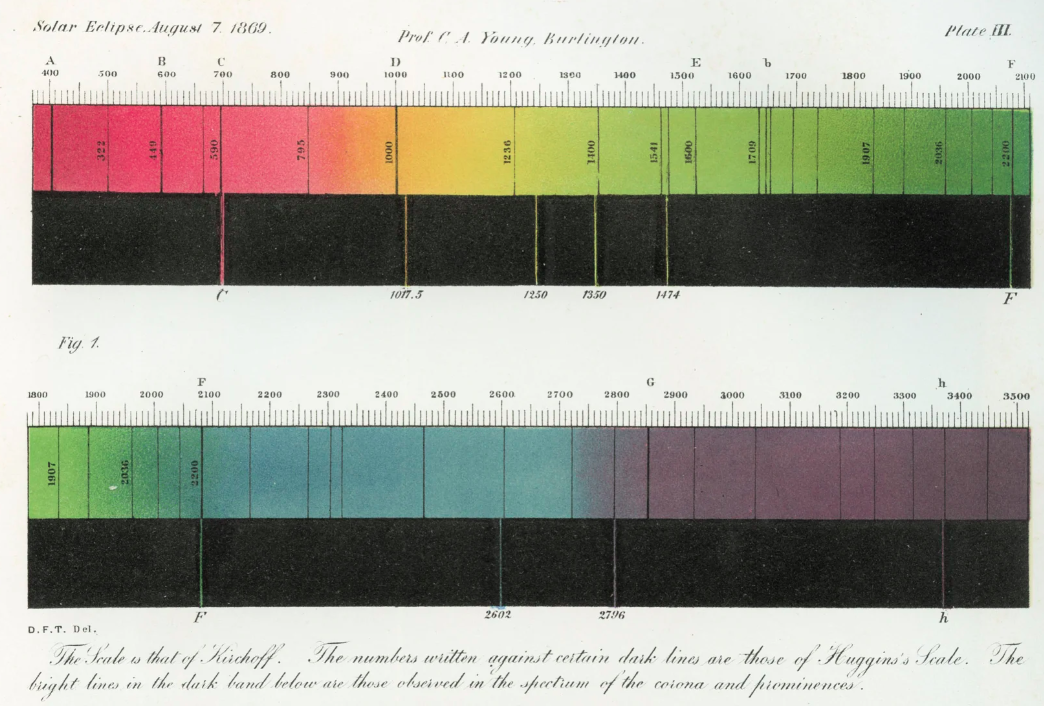}
	\caption{Spectrum observed by Charles Young during the total solar eclipse of 1869 August 7.}
	\label{fig:1}
\end{figure*}

The coronal spectroscopy began with the discovery green emission line in 1869. Charles Young made the spectroscopic observations of the solar corona using a state-of-the-art five-prism spectroscope at Burlington during the total solar eclipse of 1869 August 7. He recorded the absorption lines of the solar photosphere, the bright emission lines of chromospheric prominences, and, most remarkably, a faint green emission line from the solar corona. William Harkness also found the existence of this emission line in the corona. The green line at 5310{\AA} was a mystery as it did not correspond to any known terrestrial element. Therefore, this line was interpreted due to a new element ``coronium''. Figure \ref{fig:1} shows the spectrum observed by Young.  Later, in 1942, Edlen explained that the green line emission at 5303{\AA} is due to highly ionized iron ion. The identification of this line was an indicator of the unexpectedly high temperature of solar corona \citep{edlen1943}. Early coronal temperature estimates were based on the width of coronal emission lines. The absence of hydrogen and metal emission lines indicated that the electron temperature must be above 
7\,${\times}\,10^{5}$\,K \citep{goldberg1948}. Spectroscopy performed during the subsequent total solar eclipses helped to identify the various emission lines in visible part of the coronal spectrum. Eighteen coronal emission lines due to various ions of Fe, Ca, Ni and Ar were identified till 1937, indicating the existence of different temperatures in the coronal plasma. Low resolution spectroscopy, covering most of the visible part of the spectrum of the solar corona helps to identify the emission lines and study their abundances. High resolution spectroscopy provides information about the physical and dynamical characteristics of corona determining the intensity, width, and Doppler velocity of emission lines.  

\citet{bappu1972} carried out low resolution spectrographic observations of the solar corona during the total solar eclipse of 1970 March 7 at Mexico. They found emission lines at H${\alpha}$, He-I and H$\beta$ in the spectrum, high in the corona indicating the existence of low temperature plasma in the ambient high temperature plasma. It may be noted that some observers have found the existence of emission at H and K lines in the spectroscopy performed during the earlier eclipses having the same implication \citep{lyot1955,aly1962}. 
The slit-less evaluation of the green emission line pioneered by \citet{jarrett1955} using the Fabry-Perot etalon during the total solar eclipse of 1954 June 30 showed that decrease in line width in the corona with increasing distance from the Sun. Further, they made similar observations in the green and red emission line during the eclipse of 1958 and found that the line widths indicate temperatures of
3.2\,${\times}\,10^{6}$\,K and 
3.5\,${\times}\,10^{6}$\,K, respectively. \citet{liebenberg1974} made the spectrographic observations in the green emission line using a high resolution Fabry-Perot interferometer during the total solar eclipse of 1973 onboard Concorde aeroplane which extended the duration of the total eclipse to 74 minutes. They found periodic oscillations of $6{\pm}1$ minute in the intensity of green line but could not detect oscillations in the velocity. The green line spectroscopy using Fabry-Perot interferometer performed during the total solar eclipse of 1980 in India, and in both green and red lines during 1983 Indonesia eclipse indicated large scale motions in the corona \citep{desai1982, raju1993, chandrasekhar1991}

The Fabry-Perot instrument has the advantage to record interference fringes over most of the inner corona, but the disadvantage is the uncertainty of the contribution by the Doppler-shifted elements to the emission line profile. Such limitations are absent in the slit spectroscopy, since finite width of the slit samples limit the spatial extent of the corona. But, the slit length permits to study the corona along the length of the slit. The use of multi-slit in the spectrograph helps to study two dimensional solar corona up to certain extent \citep{livingston1970}. High resolution spectrographic observations in the red emission line during the total solar eclipse of 1980 February 16 were carried out by \citet{singh1982} using a multi-slit spectrograph. The authors found the existence of 
non-thermal velocity of 30\,km\,$\rm s^{-1}$. They did not find any systematic flows in the corona. Later using the same data, \citet{singh1985} reported that collisional excitation is predominant up to 0.2$R_{\odot}$ from the limb whereas collisional as well as radiative excitation is equally important for the coronal region between 0.2$R_{\odot}$ to 0.4$R_{\odot}$ from the limb. Radiative excitation plays a dominant role in the corona beyond 0.4$R_{\odot}$ from the limb. 

The spectrographic observations were made in five coronal emission lines due to Fe and Ni ions using the 40\,cm coronagraph at Sacramento Peak observatory by \citet{tsubaki1975, tsubaki1977}. He found oscillations in the Doppler velocity with a period of 300\,sec using the observations in the 5303{\AA} line, but no oscillations in the intensity signal. Time sequence high spectral resolution observations in the green emission line with the 40\,cm coronagraph were made by \citet{koutchmy1983}. The detailed power spectrum analysis of the data showed oscillations with periods of 300\,sec, 80\,sec and 43\,sec in the doppler velocity, but no prominent oscillations in the intensity. Spectrographic observations were carried out at the total solar eclipse on 1994 November 3 in Putre, Chile, in order to investigate the electron temperature and the expanding motion of the corona \citep{ichimoto1996}. The authors found that a coronal streamer on the eastern limb and a coronal hole on the south-pole have overall temperatures of about 1.5\,-\,1.7\,${\times}\,10^{6}$\,K and 
0.9\,-\,1.1\,${\times}\,10^{6}$\,K, respectively. \citet{ichimoto1995} made the coordinated observations of the coronal emission lines with the ground based 25\,cm coronagraph at Norikura observatory, Japan and Yohkoh/SXT. They took the spectra in [Fe xiv] green line, [Fe x] red line, and [Ca xv] yellow line representing coronal temperature of about 
2\,${\times}\,10^{6}$\,K, 1\,${\times}\,10^{6}$\,K and 
3\,${\times}\,10^{6}$\,K, respectively, with the coronagraph, and images with Yohkoh. They found the X-ray images resemble with the yellow line images that represent high temperature component of the solar corona. Further, they found that the active element of cool component, rapid changes of small structures, and localized large plasma mass motions to be co-spatial with the hot component. \citet{singh1999} obtained high resolution spectrographic observations in the [Fe xiv] 5303{\AA} and [Fe x] 6374{\AA} on number of days at several regions of the solar corona. They find that the intensity ratio of green to red line varies between 0.6\,-\,9.2 implying a temperture range of 1.2\,-\,1.6\,${\times}\,10^{6}$\,K for different coronal structures. Also, they find the width of the red line increases with height from the limb, as expected, whereas that of green line decreases with height, an unexpected result.  
\citet{harra1999} combined the observations in green line made with 25\,cm coronagraph and SoHO/CDS, and reported that the excess line broadening is caused by highly dynamical behavior in the transition region for active regions. From the comparison between edge-on loops in which the direction of magnetic field is nearly parallel to the line-of-sight direction and face-on loops in which the magnetic field is almost perpendicular to the line-of-sight direction, \citet{hara1999} found that the full width at half-maximum (FWHM, also called simply as width) of coronal emission lines for the edge-on loop appears to become smaller near the loop top than that for the face-on loops. The obvious decrease of FWHM is found in the [Fe xiv] edge-on loops. The change in velocity amplitude for this line does support the heating through Alfv{\'e}n waves, as it is too small. \citet{raju2000} combined the spectroscopic observations of the red emission line with 25\,cm coronagraph and coronal images obtained with Extreme-Ultraviolet Imaging Telescope (EIT) onboard SoHO to make temperature map of corona. They found that the typical non-thermal velocity in coronal holes is 24\,km\,$\rm s^{-1}$ while that in quiet regions is 15\,km\,$\rm s^{-1}$. The enhanced non-thermal velocity in the coronal hole is suggestive of the important role of the non-thermal broadening mechanism in the acceleration of fast solar wind.

\section{Coronal spectroscopy at present\,--\,in $21^{\mathrm{st}}$ century}

Considering the results of coronal spectroscopy done in the past during the total solar eclipses and using coronagraphs, the need was felt to do high resolution spectroscopy in multi emission lines, simultaneously. The width of the emission line, if considered as due to thermal broadening, leads to large temperature of solar corona. For example, the observed width of [Fe xiv] 5303{\AA} line yields a temperature more than 3\,${\times}\,10^{6}$\,K whereas, the abundance of ions, emitting [Fe xiv] line, indicates temperature of corona to be about 
2\,${\times}\,10^{6}$\,K. Also, other observations such as scale-height temperature indicate that temperature of the corona to be 1\,-\,2\,${\times}\,10^{6}$\,K. Hence, the excess line width of the coronal emission lines has been interpreted as due to non-thermal component. Generally, scientists assume the temperature due to maximum abundance of the ion depending on the theoretical models as thermal component and then derive the non-thermal component of the solar corona. But, in reality, different structures of the solar corona have different temperatures. Therefore, it is not possible to separate out the two components from a single emission line observation, indicating the need of multi line spectroscopy, simultaneously.

The solar corona is inhomogeneous, highly structured, and dynamic. The temperature varies over the range 1\,-\,3\,${\times}\,10^{6}$\,K in quiet coronal structures, and 2\,-\,10\,${\times}\,10^{6}$\,K during active events. The abundances of different ions vary depending of the temperature of the structure, thereby, the intensity of the emission line. Therefore, spectroscopic observations are required in multiple emission lines representing these temperatures. Emission lines due to different ions of Fe species in the visible part of the spectrum represent 1\,-\,2\,${\times}\,10^{6}$\,K temperature. The yellow emission line due to [Ca xv] represents temperature around 
3\,${\times}\,10^{6}$\,K and is weak in the quiet regions of the solar corona.

\subsection{Research with 25\,cm coronagraph in the Norikura Observatory, Japan}

\begin{figure*}[!ht]
	\centering
	\includegraphics[width=0.95\linewidth]{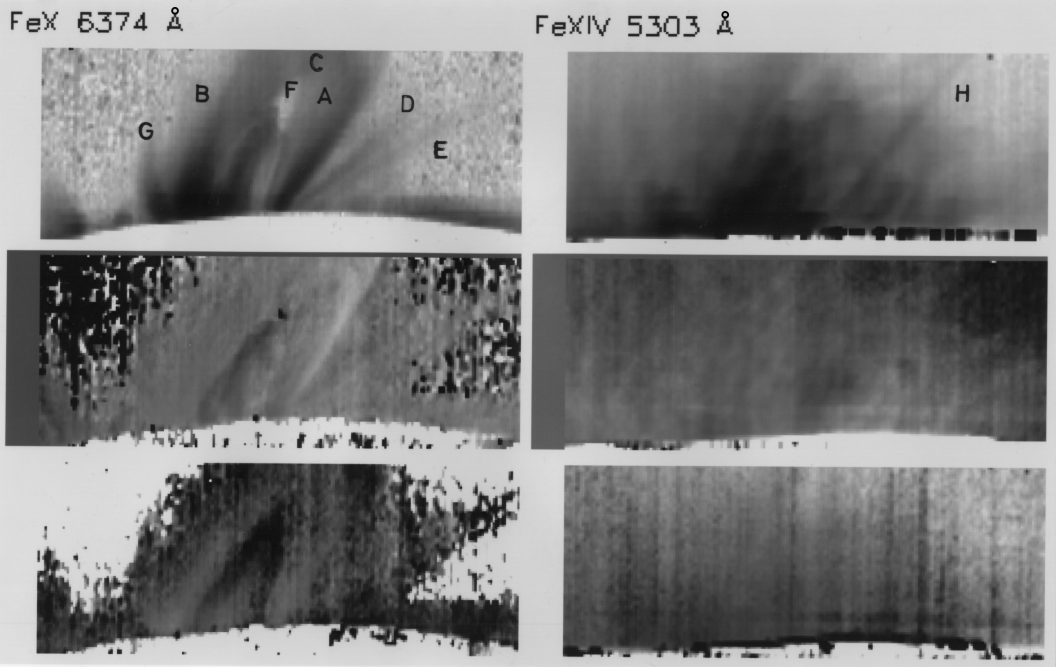}
	\caption{This figure is adopted from the paper \citep{singh2002}. Top two panels show the intensity of the coronal region in the red and green lines. Middle and bottom panel show the corresponding FWHM and the Doppler velocity.}
	\label{fig:2}
\end{figure*}

In view of the above, \citet{singh1999, singh2002, singh2003a, singh2004, singh2006} made the systematic spectrographic observations in [Fe x] 6374{\AA}, [Fe xi] 7892{\AA}, [Fe xiii] 10747{\AA} \& 10798{\AA}, and [Fe xiv] 5303{\AA} lines, choosing 3 or 4 out of these five emission lines, simultaneously during the period of 1997\,-\,2007. Simultaneous, spectroscopic observations made for the first time using 2 CCD cameras. Such observations have the advantage of comparing the parameters of different emissions lines at the same time and at the same location in the corona without any ambiguity. In the following we discuss the various results from the systematic observations with 25\,cm coronagraph.  Figure \ref{fig:2} shows a coronal region observed on 1998 July 19 in the red and green line, simultaneously.  Top two panels show the intensity of the coronal structures in the red and green line. Middle and bottom panels show the corresponding FWHM and Doppler velocity in the red and green line.

The simultaneous observations in the green and red emission lines, indicate that the FWHM of the red line increases with height above the limb at a rate of 
0.5\,-\,2.6 m{\AA}\,$\rm arcsec^{-1}$ and width of the green line decreases with height at a rate of 
1.2\,-\,3.0 m{\AA}\,$\rm arcsec^{-1}$ \citep{singh2002}. The difference in the widths of the green line in open and closed coronal structures at a given height above the limb is small as compared to that between the open and closed loop structures for the red line. Further, observations in 10747{\AA} and 10798{\AA} [Fe xiii] lines on other days indicate that the ratio of the intensities of the these emission lines in the individual coronal structures range between 1.0 and 2.5 at 10 arcsec above the solar limb, which corresponds to a density range of 9.8\,${\times}\,10^{9}$\,-\,
2.4\,${\times}\,10^{8}$\,$\rm cm^{-3}$ \citep{singh2002}. The scale-height temperature values, derived from the variations of the intensity ratio with height above the limb for all individual coronal structures, range between 0.6 $\times$ $10^{6}$\,K and 8.3 $\times$ $10^{6}$ K with a most frequent value around 1.8 $\times$ $10^{6}$ K. The large values of the scale-height temperature for 70\% of the structures indicate that these structures may be neither in hydrostatic equilibrium nor isothermal in nature. Time sequence for 80\,min with a cadence of 25\,sec was made of coronal green-line spectra at a fixed location in the corona with 25\,cm coronagraph \citep{sakurai2002}. The Fourier analysis of Doppler velocity derived from the profiles shows powers in the 1\,-\,3\,mHz range, and in higher frequencies (5\,-\,7\,mHz) at localized regions. The line intensity and line width do not show clear oscillations, but their phase relationship with the Doppler velocity indicates propagating waves rather than standing waves. They found that the Alfv{\'e}n waves whose speed is 500\,km\,$\rm s^{-1}$ or faster are possible but inconclusive, while the existence of slower waves (of the order of 100\,km\,$\rm s^{-1}$, possibly sound waves) is evident. But, the energy carried by the detected sound waves is far smaller than the required heat input rate to the quiet corona. Spectra around the 6374{\AA} [Fe x] and 7892{\AA} [Fe xi] emission lines were obtained in a raster scan mode \citep{singh2003a}. The line width (FWHM) of the 6374{\AA} emission line increases with height above the limb with an average value of 
1.02\,m{\AA}\,$\rm arcsec^{-1}$ in steady coronal structures. The FWHM of the 7892{\AA} line also increases with height but at a smaller value of 0.55\,m{\AA}\,$\rm arcsec^{-1}$. The result is against the expectation, a similar increase in non-thermal or temperature in both these lines would have shown larger gradient for 7892{\AA} lines as compared to that for 6374{\AA} line but observations indicate opposite behavior.

\begin{figure*}[!ht]
	\centering
	\includegraphics[width=0.95\linewidth]{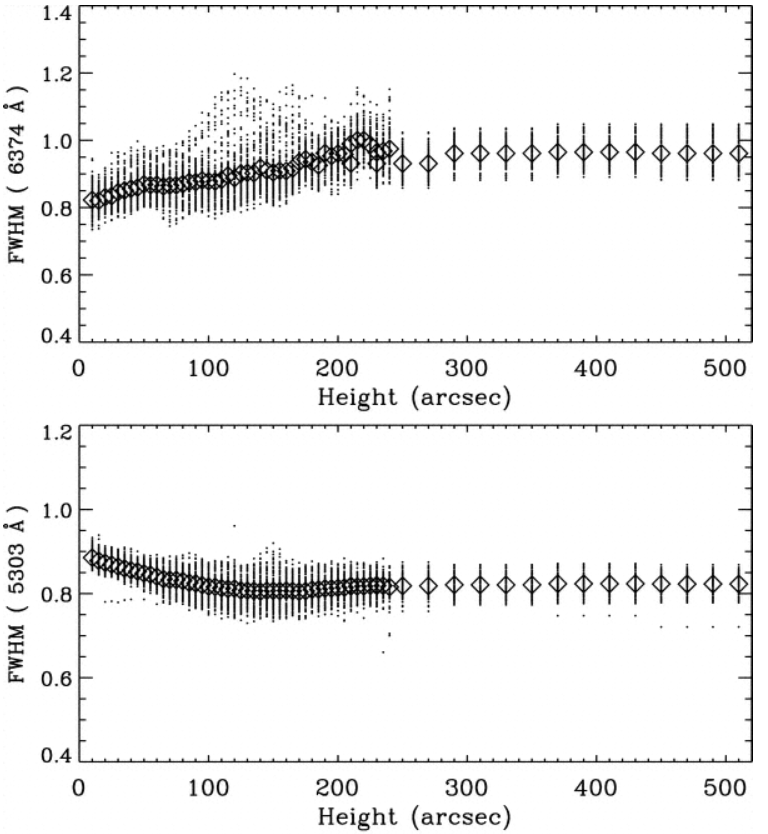}
	\caption{Variation of FWHM of the red and green line as  a function of height above the solar limb indicates that the FWHM remains the same after about 250\,arcsec.}
	\label{fig:3}
\end{figure*}

\citet{singh2003b} investigated the variations in the FWHM of [Fe xiv] 5303{\AA}, [Fe x] 6374{\AA}, [Fe xiii] 10747{\AA} and [Fe xiii] 10798{\AA} emission lines in number of coronal structures on several days. The FWHM of the emission lines increases at an average rate of 1.24\,m{\AA}\,$\rm arcsec^{-1}$ for [Fe x], 0.29\,m{\AA}\,$\rm arcsec^{-1}$ for [Fe xiii], and decreases at the rate of - 0.66 m{\AA}\,$\rm arcsec^{-1}$ for [Fe xiv]. These values indicate an inverse correlation with the corresponding ionization temperature for these emission lines. They inferred that FWHM of emission lines increases with height if the associated ionization temperature is less than 
1.6 $\times$ $10^{6}$ K, with the gradient depending upon the ionization temperature of the line. It decreases with height for lines whose ionization temperature is greater than 1.6 $\times$ $10^{6}$ K. It implies that it is not possible to interpret the observed increase in FWHM with height in terms of an increase in non-thermal velocity. Spectroscopic observations obtained simultaneously in pairs of coronal emission lines, one line being [Fe x] 6374{\AA} and the other line being [Fe xi] 7892{\AA}, [Fe xiii] 10747{\AA}, or [Fe xiv] 5303{\AA}, were analyzed to investigate the variations in the intensity and FWHM ratios of these lines with respect to those of 6374{\AA} as a function of height above the limb \citep{singh2004}. The intensity ratio of the 7892{\AA} and 10747{\AA} lines with respect to the 6374{\AA} line increased with height and that of 5303{\AA} to 6374{\AA} decreased with height above the limb. The results imply that the temperature in coronal loops will increase with height if one considers the intensity ratio of 7892{\AA} to 6374{\AA} line and will appear to decrease with if height one considers the intensity ratio of 5303{\AA} to 6374{\AA}. In addition, the varying FWHM ratios of these lines with height showed that hotter and colder plasmas in coronal loops mix with each other. 
Time-sequence spectroscopic observations in [Fe x] 6374{\AA} and [Fe xiv] 5303{\AA} lines with the 25\,cm coronagraph, and narrow-band and Doppler images in 5303{\AA} line with 10\,cm Doppler coronagraph ``NOGIS'' at the Norikura Observatory, of a coronal region for about 7\,hrs were analyzed \citep{singh2005}. The coronal region showed the formation of a coronal loop by a high-speed surge in the 6374{\AA} line rising from one of the foot points of the loop. The spectroscopy in the 6374{\AA} line showed large velocities along the line of sight and vertical to the solar limb whereas the 5303{\AA} line observations indicated negligible line-of-sight velocities and low vertical velocities as compared to those in the 6374{\AA} line. The low values of FWHM at the beginning of jet like structure suggest that a relatively cooler plasma at a temperature of about 0.7\,${\times}\,10^{6}$\,K or less was ejected from the transition region with a large velocity and heated up in the corona by some process and formed a coronal loop. These observations indicate that the loop was formed by evaporation of chromospheric plasma. 
With a view to determine the variations in the parameters of coronal emission lines up to about 500\,arcsec above the solar limb, \citet{singh2006} performed spectroscopy in raster scans mode with large exposure times in the range of 30 to 200\,sec for spectra, exposure time varying with distance from limb. They found that the FWHM of the [Fe xiv] 5303{\AA}  line decreases up to about 300\,{$\pm$}\,50\,arcsec and then remains more or less the same up to 500\,arcsec, while that of the [Fe x] 6374{\AA} line increases up to about 250\,arcsec and then remains unchanged. The FWHM of the [Fe xi] 7892{\AA} and [Fe xiii] 10747{\AA} lines show an intermediate behavior. Figure \ref{fig:3} shows the variation of FWHM in green and red lines as a function of height above the limb. The non variability in the FWHM of emission lines after about 300\,arcsec above the limb in steady coronal structures does not support the prevailing view that the non-thermal velocity increases with height due to either the coronal waves or the high-velocity solar wind. 
 
\begin{figure*}[!ht]
	\centering
	\includegraphics[width=0.95\linewidth]{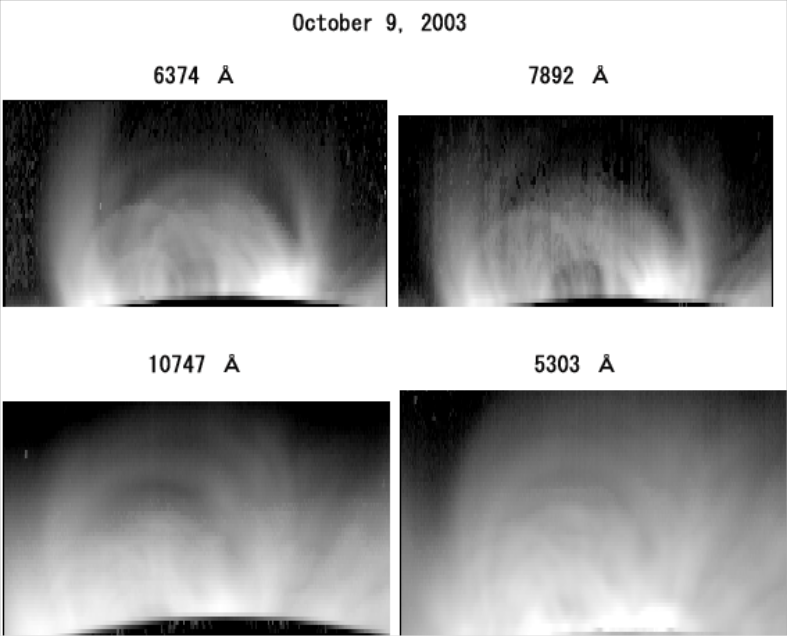}
	\caption{Four panels of the figure show the same coronal region observed in four emission lines representing temperature in the range of 1\,-\,2\,${\times}\,10^{6}$\,K.}
	\label{fig:4}
\end{figure*}

\begin{figure*}[!ht]
	\centering
	\includegraphics[width=0.95\linewidth]{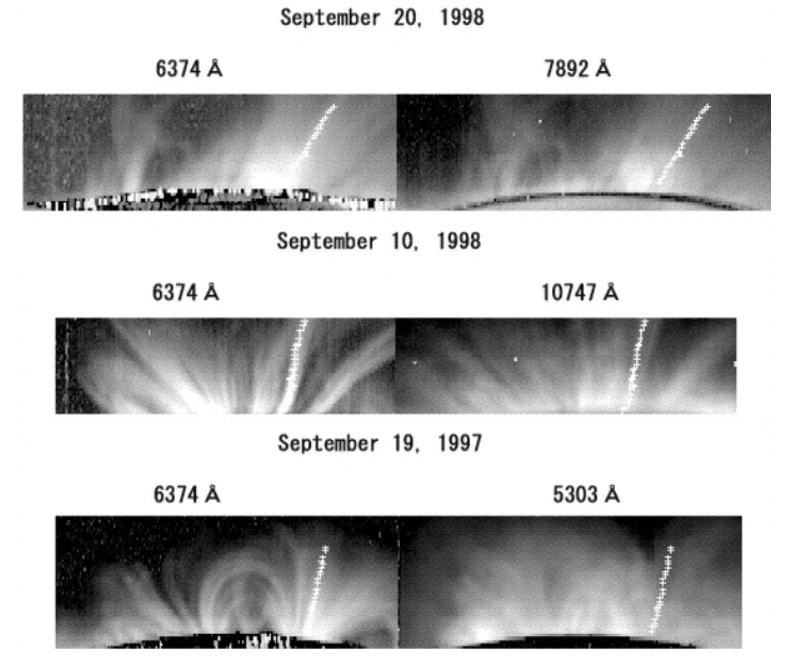}
	\caption{The intensity images in 6374{\AA} and the corresponding images in 7892{\AA}, 10747{\AA} and 5303{\AA} emission lines for different coronal regions obtained on different days used to investigate the intensity ratios. The `+' marks indicate the loops used for the study.}
	\label{fig:5}
\end{figure*}

\begin{figure*}[!ht]
	\centering
	\includegraphics[width=0.95\linewidth]{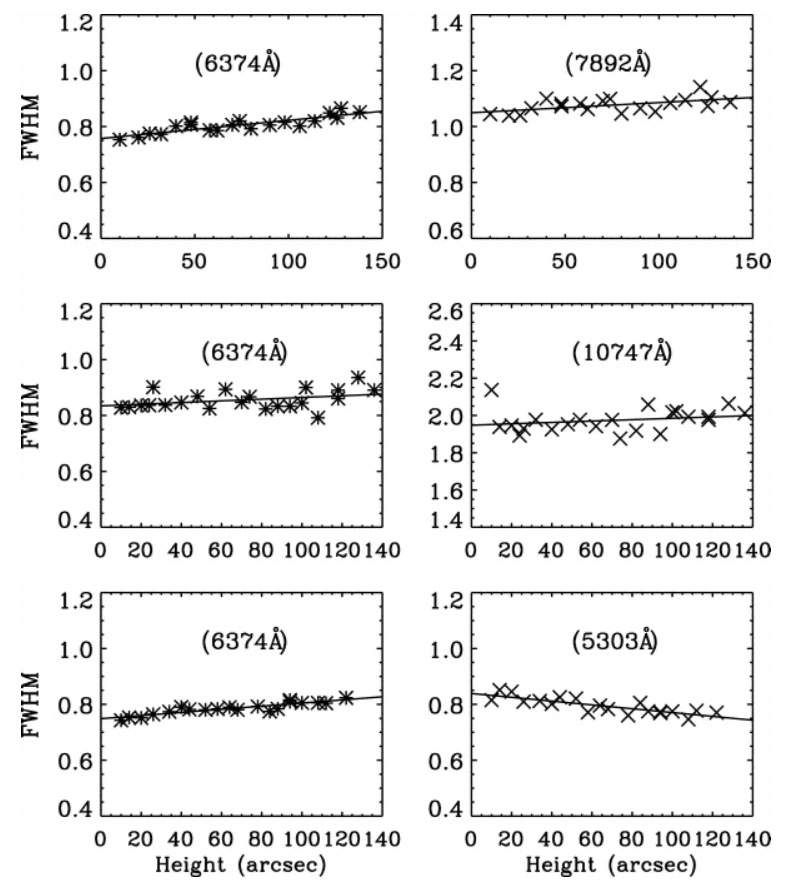}
	\caption{FWHM of 6374{\AA} line and that of other three emission lines for the loops indicated in Figure 5.}
	\label{fig:6}
\end{figure*}

\begin{figure*}[!ht]
	\centering
	\includegraphics[width=0.95\linewidth]{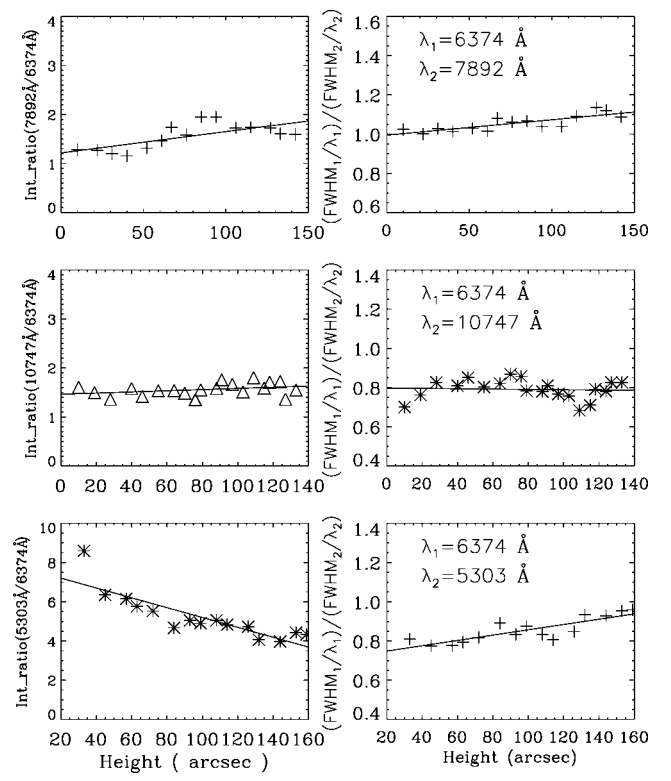}
	\caption{Intensity and FWHM ratios of the 7892{\AA}, 10747{\AA} and 5303{\AA} emission lines with respect to 6374{\AA} line.}
	\label{fig:7}
\end{figure*}

In the following we summarize the results of all the systematic results and their implications:

Figure \ref{fig:4} shows the same coronal region observed in four emission lines. The [Fe x] and [Fe xi] intensity images represent relatively low temperature around 
1\,${\times}\,10^{6}$ K whereas [Fe xiii] and [Fe xiv] represents temperature relatively high around 
2\,${\times}\,10^{6}$ K. The structures seen in low temperature lines are sharper as compared to those in high temperature lines.
Figure \ref{fig:5} shows the intensity images in 6374{\AA} and the corresponding images in 7892{\AA}, 10747{\AA} and 5303{\AA} emission lines for different coronal regions obtained on different days used to investigate the intensity ratios as a function of height above the limb. The loops were selected in the 6374{\AA} images shown by `+' marks and the corresponding locations in other line images were automatically selected. Figure \ref{fig:6} shows the variation of FWHM for the selected coronal loop as a function of height.
Figure \ref{fig:7} indicates the intensity and FWHM ratios of the 7892{\AA}, 10747{\AA} and 5303{\AA} emission lines with respect to 6374{\AA} line. The line widths have been normalized by dividing these by respective wavelength while computing the ratios.

\subsubsection{Summary of the results}

\begin{figure*}[!ht]
	\centering
	\includegraphics[width=0.95\linewidth]{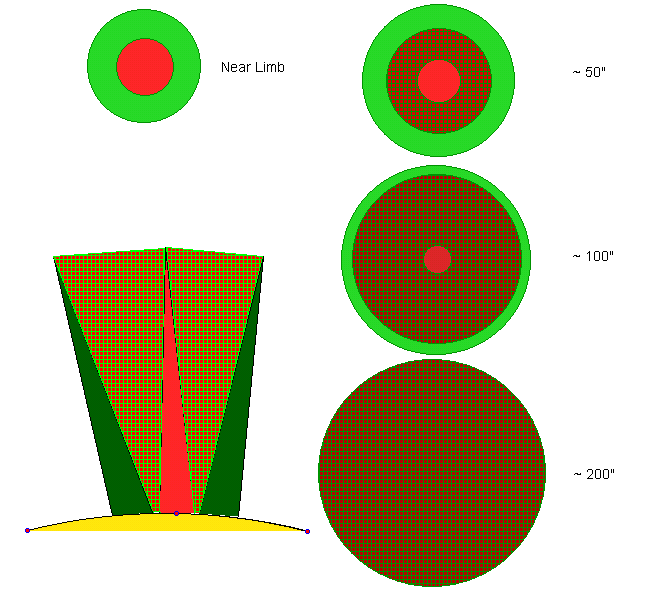}
	\caption{Cartoon showing the mixing of red (cooler) and green (hotter) lines plasma at different heights.}
	\label{fig:8}
\end{figure*}

\begin{itemize}

\item All these observations indicate that variation in line width with height is related with the temperature of the ion which the line represents.

\item Emission lines associated with temperature of maximum abundances greater than 1.6\,${\times}\,10^{6}$\,K show a decrease in line width with height, whereas those associated with temperatures less than 1.6\,${\times}\,10^{6}$\,K show an increase. The slope of variation depends on the associated temperature of emission line and is negligible for emission line around 1.6\,${\times}\,10^{6}$\,K.


\item The trend in the variation of line width is independent of the shape, size, topology of the coronal loops. To explain it further, line width of the 5303{\AA} line decreases in all types of loops such as small or big$;$ open or closed$;$ face-on or edge-on$;$ radial or non-radial and that of 6374{\AA} increases with height. The magnitude of variation may be different for different coronal loops, because it is likely to depend upon the underlying magnetic field, density, temperature etc.

\item The FWHM of [Fe x] and [Fe xiv] emission lines do not show change with height after a distance of about 250\,arcsec and normalized ratio is 1, suggesting that plasma at larger heights are at uniform temperature and non-thermal velocities. The distance of 250\,arcsec may depend on the physical properties of coronal structure.

\item The FWHM gradient is negative for high-temperature emission lines, positive for relatively low-temperature lines and smaller for emission lines in the intermediate temperature range. Such behavior in the variation of the FWHM of coronal emission lines with height above the limb indicate that increase or decrease in non-thermal velocity with height alone may not fully explain the observed behavior. \citet{hahn2014} determined the energy and dissipation of Alfv{\'e}nic waves in the quiet Sun using spectral lines to infer the wave amplitude. They find that waves cause non-thermal broadening of the spectral lines, which can be expressed as a non-thermal velocity. They also report that these waves get dissipated over a region centred on the top of the loops. The excess line width of emission lines over the thermal broadening is an observational constraint for both nano-flare and wave dissipation models of coronal heating \citep{brooks2016}. From observations of non-flaring active regions using Hinode Extreme Ultraviolet Imaging Spectrometer (EIS) they find relatively small non-thermal velocities with a mean value of 17\,km\,$\rm s^{-1}$. There is no significant trend with temperature or active region magnetic flux. They infer that these measurements appear to be inconsistent with those expected of  reconnection jets in the corona, chromospheric evaporation induced by coronal nano-flares, and Alfv{\'e}n wave turbulence models. \citet{delzanna2018} discussed in detail the role of non-thermal velocity and waves in their review of observations of the outer solar atmosphere of the Sun, especially of the corona, made in visible, UV, EUV and X-ray wavelength range. \citet{sudheer2023} studied a blowout jet observed by the Solar Dynamics Observatory$/$Atmospheric Imaging Assembly and the Interface Region Imaging Spectrograph. They found that line width showed periodic oscillations with a period of $\approx$3 minutes, and plasma upflow always occurred when the line width was large. They argued that multiple reconnections might produce periodic line broadening. Further, they conclude that oscillations in emission line intensity and width indicate multiple magnetic reconnection triggers quasi-periodic fluctuations (QPPs) in the blowout jet. Recently, \citet{Ramesh2026b} reported energy releases during pre-flare activity leading to simultaneous line broadening and plasma upflows using observations a coronal loop with the VELC. The estimated upper
limit for the Alfv{\' e}n wave energy flux at $r{\approx}1.1R_{\odot}$ in the loop is ${\approx}$39\%.
\citet{srivastava2025} suggest that the dissipation of magnetic waves and direct electric currents are likely to be the significant processes for heating the corona because magnetic waves may trigger current dissipation, and impulsive reconnection can launch magnetic waves. They present numerical simulations to support the concept of symbiosis between waves and reconnection in heating the solar corona.

\item The intensity ratio of [Fe xi] to [Fe x] line generally increases with height above the limb indicating increase in temperature with height. This agrees with the findings of \citet{kano1995} showing that emission measure is highest at the loop top.

\item The intensity ratio of [Fe xiv] to [Fe x] line decreases with height above the limb, an unexpected result. The decrease in this intensity ratio implies the loop top is cooler if observed in [Fe xiv] emission line.

\item The present results indicate the inadequacy of the earlier coronal loop models. The monotonic increase in temperature or non-thermal velocity with height above the limb cannot explain the observed variations in line widths and intensity ratios with height in all the lines, simultaneously. The observed variations can be explained if we assume that the thin coronal loops are not magnetically shielded. The different temperature plasma interacts with each other, probably due to collision and the whole of plasma attains uniform temperature and non-thermal velocity at larger heights as explained in Figure \ref{fig:8}. This is an empirical model and needs to be confirmed with more observations. The process of collision between ions in the low density plasma at high temperatures needs to be investigated. This scenario is supported by the model of coronal loops proposed by \citet{akiyama2005}. They propose that the different temperature regions are not magnetically isolated from each other and can therefore be physically connected through thermal conduction.

\item One of the scenarios can be spontaneous heating of plasma by small scale reconnections occurring at the base of the loops producing different temperature plasma depending on the energy involved in the reconnection. Then different temperature plasma interacts with each other creating a uniform plasma at higher heights in the corona.

\end{itemize}

\subsection{Other similar investigations}

LASCO-C1 onboard SoHO had the capability to perform spectroscopy in the visible wavelength band in the solar corona above 1.1$R_{\odot}$. Using series of images acquired at different wavelengths across the Fe x 6374{\AA} (red) and Fe xiv 5303{\AA} (green) coronal lines, \citet{mierla2008} made radiance and line width maps of the off-limb solar corona. They found that the width of [Fe xiv] emission line remained same or increased with height up to around 1.3$R_{\odot}$ and then it decreased. The [Fe x] line width increased with height up to about 1.3$R_{\odot}$. [Fe x] line showed higher effective temperatures as compared with [Fe xiv]. It may be noted that instrumental profile has a width of 0.7{\AA} comparable with the average width of of the green and red lines. 
\citet{contesse2004} observed the solar corona between 1.6\,-\,8\,arcmin above the limb in the four directions (E, W, N, \& S) using a 40\,cm aperture coronagraph. They found that the line widths and turbulent (non-thermal) velocities are significantly higher above the polar regions, especially during a coronal hole. They did not find any decreasing behavior of widths and turbulent velocities in equatorial directions for larger radial distances. The variation in the high corona is rather flat which agrees with the findings of \citet{singh2006} which indicates that the average widths of red and green line are 0.95{\AA} and 0.82{\AA} at 250\,arcsec above the limb that remains same at higher heights. \citet{singh2006} report the decrease in width with height up to about 200\,arcsec only. 
The thermal structure of a coronal loop, both along and across the loop, is vital in determining the exact plasma heating mechanism. Observations on a number of days were made, simultaneously in four forbidden iron emission lines namely, [Fe xi] 7892{\AA} line, [Fe xiii] 10747{\AA} \& 10798{\AA} lines and [Fe xiv] 5303{\AA} line and on some days made only in [Fe xi] 7892{\AA} and [Fe x] 6374{\AA} lines \citep{prasad2013}. The intensity ratios of different lines indicated a significant negative temperature gradient in all the structures observed in [Fe xiv] and [Fe xiii] and a positive temperature gradient in the structures observed in [Fe xi] and [Fe x]. This implies that the loop tops appear hotter when observed in relatively colder lines and colder when observed in relatively hotter lines as compared to their coronal foot points. This contrasting trend observed in the temperature variation along the loop structures can be explained by a gradual interaction of different temperature plasma. 
\citet{voulgaris2010, voulgaris2012} obtained the slit-less flash spectra during the eclipses of 2006 and 2008.  The measured intensity of the 5303{\AA} emission line indicated that the ionization state of the corona, as manifested especially by the [Fe xiv] emission line, was much weaker during the 2008 eclipse as compared to 2006 eclipse, indicating that following the long, inactive period during the solar minimum, there was a drop in the overall temperature of the solar corona. \citet{voulgaris2012} found the intensity ratio of green to red line during the 2010 eclipse is different than those during the 2008 eclipse. They could also found the existence of weak emission in [Ca xv] line. 

During the totality phase of 2017 eclipse, \citet{koutchmy2019} took the spectra using a long slit, covering $\pm$3$R_{\odot}$ in the range of 5100{\AA} to 5900{\AA}. They detected rarely observed [Ar x] line at most of the locations in the corona along with the [Fe xiv] and [Ni xiii] lines. They could also observe hot lines inside the coronal holes regions. They did not find large Doppler shifts anywhere in the inner or the middle corona. \citet{Habbal2011} demonstrated the unique diagnostic capabilities of the coronal forbidden lines for exploring the evolution of the coronal magnetic field and the thermodynamics of the coronal plasma, using data obtained during the 2010 July 11 eclipse observations. The advantages of mutli-site observations for probing the dynamic and thermodynamic properties of the corona, and the impact of transients like the CMEs, were reported by \citet{Boe2020} during the 2017 August 21 eclipse. \citet{Zhu2024} showed that observations during the eclipses provide opportunity to measure the line widths simultaneously in open and closed ﬁelds at larger heights above the solar limb, with minimal exposure times, stray light levels, and instrumental widths.   

Currently, the 4\,m aperture Daniel K. Inouye Solar Telescope (DKIST) provides the highest-resolution observations of the Sun ever achieved. The design of the telescope permits observations of the solar corona at infrared wavelengths where the sky background is low, and bright coronal emission lines are available \citep{kuhn1996, judge2002, lin2004, dima2019, rimmele2020, schad2024a}. The large aperture of DKIST combined with state-of-the-art instrumentation permits to measure the vector magnetic field in the chromosphere and in the faint corona with required sensitivity. 
Mapping of the coronal magnetic field is a key objective of the DKIST to understand various aspects of energetic events, space weather and to improve prediction capabilities. 
Recently, \citet{schad2024b} reported DKIST maps of polarized spectra of the active region corona in the 10747{\AA} [Fe xiii] emission line. Clear signatures of Zeeman effect due to the coronal magnetic field is seen.
Their comparisons with global magnetohydrodynamic models provide valuable constraints for coronal modeling efforts. Observations of the global coronal magnetic field similar to that reported by \citet{Yang2024} using the ground based Upgraded Coronal Multi-channel Polarimter (UCoMP) in the 1074\,nm [Fe xiii] emission line are also expected to be useful in this connection.

\subsection{Visible Emission Line Coronagraph (VELC) onboard Aditya-L1}

\citet{singh2006} planned observation for about 600 days with Norikura 25\,cm coronagraph during the period 1997- 2007 but could get the data only for about 100 days due to the sky conditions.  Also, they could get the data mostly for couple of hours on each day except for 6\,-\,7 days when data could be obtained more than 6\,hrs a day. It is important to get  continuous data without the interruption of day-night cycle in multiple emission line as is evident from the data obtained from VELC described in this section. \citet{singh2025} had planned to perform the coronal spectroscopy in 4 bright emission lines in the visible part of the spectrum from an instrument in space to begin with. Due to restriction on the size and weight of satellite, it was planned to do so in 3 lines, namely, [Fe xi] 7892{\AA}, [Fe xiii] 10747{\AA} and [Fe xiv] 5303{\AA}, simultaneously along with the imaging in broad band. Now, the data is available for long periods for each day, during most of the days in a year in the [Fe xiv] emission line only. The scattered light in the instrument does not permit to get meaningful results from the other two relatively weak emission lines. The VELC onboard Aditya-L1 was launched in 2023 and is used to make observations in sit-and-stare and raster scan mode, since then.

\begin{figure*}[!ht]
	\centering
	\includegraphics[width=0.95\linewidth]{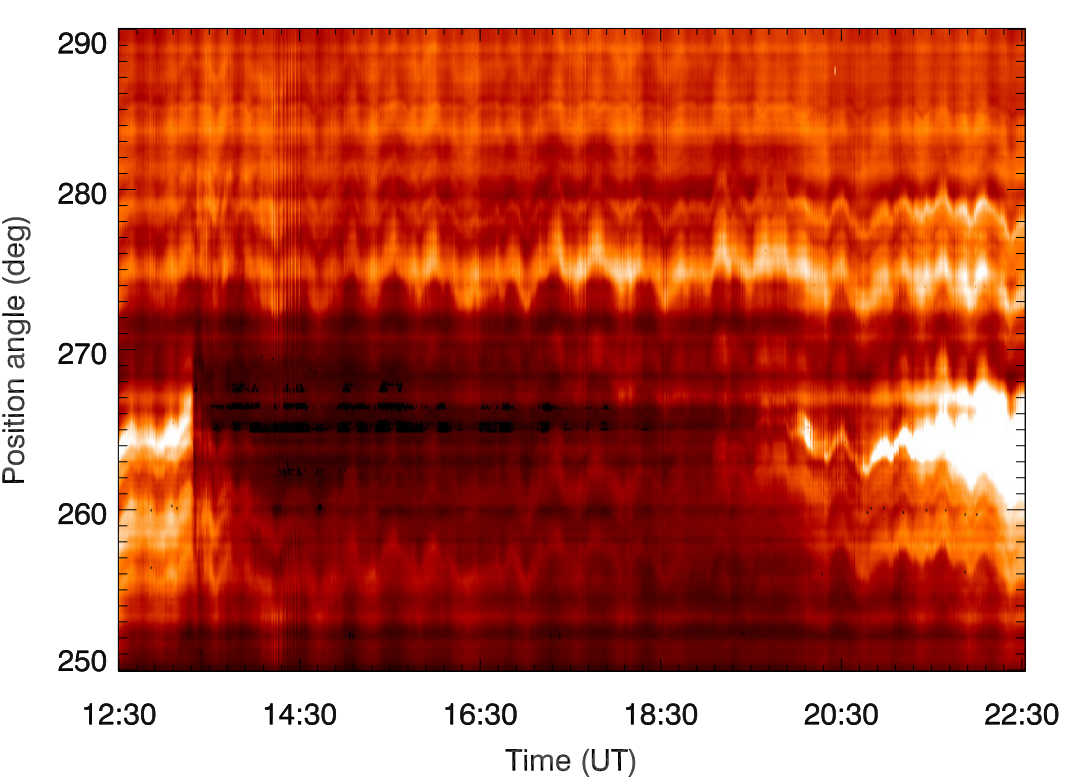}
	\caption{Variations in the peak intensities of the 5303{\AA} emission line in the solar corona  at 1.05$R_{\odot}$ observed on 2024 July 16 July with Aditya-L1/VELC. A sudden dimming of the intensity can be clearly observed in the position angle range 
${\approx}260^{\circ}$\,-\,$270^{\circ}$ during the interval 
${\approx}$13:18\,-\,19:00\,UT. The reasons for the oscillatory pattern in the image may be due to drift in the satellite causing shift in the image.}
	\label{fig:9}
\end{figure*}

A coronal mass ejection (CME) was observed in the corona towards the west side of Sun and an X1.9 class soft X-ray flare also occurred near the west limb on 2024 July 16 \citep{ramesh2024}.  Figure \ref{fig:9} shows variations in the peak intensities of the green line (5303{\AA}) emission line in the limited region (by slit) of solar corona at 1.05$R_{\odot}$ as a function of time,  observed on 16 July 2024 with VELC onboard Aditya-L1. A sudden dimming of the intensity is seen at between the position angle ${\approx}260^{\circ}$\,-\,$270^{\circ}$ during the interval 
${\approx}$13:18\,-\,19:00\,UT. The oscillatory pattern seen in the image is likely due to drift in the satellite causing shift in the image. After the dimming in the region again green line emission shows the brightening in the same region, probably due to the another event. During the same time spectroscopic observations were made with the VELC in 5303{\AA} emission line at 1.05$R_{\odot}$ on the west limb. From the VELC observations they could compute the more reliable onset time of the CME. They found ${\approx}$50\% decrease in the coronal intensity near the source region of the CME due to mass depletion, ${\approx}$15\% enhancement in the emission line width, and enhancement in Doppler velocity of 
${\approx}$10\,km\,$\rm s^{-1}$. The computed non-thermal velocity from the line broadening is ${\approx}$24.87\,km$\rm s^{-1}$.  \citet{priyal2025a} made long-duration observations with VELC of activity associated with a flare-less CME erupted from an active region at the west limb of the Sun. The spectroscopic observations were in the sit-and-stare mode for about 7\,hrs. The analysis showed a steady increase in the intensity of the line by 
about ${\approx}$57\%, followed by a gradual decrease to the initial level. The variations in line width were anti-correlated with those in intensity. 

For another event of CME in the solar corona, \citet{priyal2025b} from the observations in 5303{\AA} [Fe xiv] estimated the electron density (Ne), thermal energy ($E_{CME}$), and mass ($M_{CME}$) of a coronal mass ejection (CME) very close to the Sun at heliocentric distance $r{\approx}1.06R_{\odot}$. Using spectroscopic observations in the 5303{\AA} coronal emission line with the VELC, \citet{Ramesh2026a} estimated the Alfv{\'e}n wave energy flux before and during a transient coronal dimming caused by a CME. The analysis indicated that the coronal energy flux associated with the dimming is ${\approx}$7\% of the typical energy loss from an active region due to conduction, radiative, and solar wind fluxes. 

\citet{Agarwal2026} carried out analysis of imaging observations with PROBA-3/ASPIICS and spectroscopic observations with VELC of two coronal mass ejections (CMEs) in the low corona, both in the 5303{\AA} emission line. They estimated the CME mass, volume, number density, and the evolution of kinetic, thermal, and magnetic energies. VELC [Fe xiv] 5303{\AA} observations provide independent estimates of the emission measure, electron number density, thermal energy, and CME lateral extent at 
${\approx}$1.06$R_{\odot}$. The two events exhibit distinctly different energy partitioning in the low corona. For the 2025 September 14 CME, the kinetic and magnetic energies are comparable, while the thermal energy remains nearly two orders of magnitude smaller, indicating limited plasma heating. In contrast, the 2025 September 16 CME shows thermal energy enhancement and eventually becomes comparable to the kinetic energy, suggesting substantial plasma heating in the low corona. Their results highlight that observational synergy between ASPIICS and VELC has the potential to constrain CME energy and density evolution in the low corona.

One of the debated issues about the onset of the type II radio bursts near the Sun is the heliocentric distance at which the associated magnetohydrodynamic (MHD) shocks are formed, and the association of the latter with the coronal mass ejections (CMEs). The debate is primarily due to the absence of routine CME observations in white light close to the solar limb. Joint observations with the VELC and Gauribidanur radio facilities \citep{ramesh2012} could be useful to address the issue as shown in \citet{kathiravan2026}.

\section{Coronal spectroscopy in future}

The VELC was planned to make the spectroscopic observations in three coronal emission lines, simultaneously, for long periods during the day and large number of days during a year restricted by the technical reasons only, to understand the physical and dynamical nature of the solar corona. But, the data in one emission line (5303{\AA}) is useful. The data obtained for long periods has yielded new scientific results and shown the versatility of such type of observations. 

\citet{Agarwal2026} have shown that the combined data of green line imaging with PROBA-3$/$ASPIICS and spectroscopic observations with VELC can help to understand the various features of the CMEs. Similarly, \citet{kathiravan2026,Ramesh2026a} have shown that combined study of the radio and VELC spectroscopic data can help in understanding the generation of shock waves and coronal dimming due to CME$’$s, respectively. The available UV, EUV, X-ray imaging and spectroscopic data from various space based instruments, data in the radio wavelength region from number of ground based observational facilities, and imaging data about flares in H${\alpha}$ and other wavelengths in the visible region can be studied simultaneously to understand the occurrence of energetic events in detail which will help in modeling of these events and heating of the coronal plasma. 
VELC data recorded for longer duration and with shorter cadence will help in making the detailed picture of the events studied. \citet{Zhangx2022} showed that 
among the correlation coefﬁcients obtained between the intensities of 5303{\AA} and the different EUV lines, the coefﬁcients
between emission in the 5303{\AA} and the 211{\AA} lines for different coronal structures and different limb locations are always good, ranging from 0.89 to 0.99. The 5303{\AA} observations are with the 10\,cm coronagraph at the Lijiang Station of the Yunnan Observatories in China. There are also interesting observations of coronal cavities above prominences using the 20\,cm coronagraph in the same place and the Atmospheric Imaging Assembly (AIA, \citealp{Lemen2012}) onboard the Solar Dynamics Observatory (SDO). The simultaneous observations indicate that the cavities are seen in
5303{\AA} and 193{\AA} emission lines, but not in 6374{\AA} and 171{\AA} \citep{Huangfu2025}. Likewise, observations of a CME front with inclusions of cool prominence material were reported earlier by \citet{Ding2017} from 2015 March 20 total solar eclipse observations in the 5303{\AA} emission line. According to \citet{Landi2016}, ground  based visible observations of the solar corona provide opportunities to address open problems in
solar physics, such as coronal heating, solar wind acceleration, and CME triggering and evolution to name a few. Combined study of magnetic field measurements with DKIST and spectroscopic data from VELC for long periods before, during, and after the energetic events can help to understand the triggering phenomena of these events.

Future space missions needs to be planned to obtain such observations in four emission lines representing temperatures between 
1\,-\,2\,${\times}\,10^{6}$\,K. The following type of experiments may be planned in future.

\begin{itemize}

\item Continuous high resolution spectroscopic observations are required with space based instruments as observations with ground based instruments are limited due to sky conditions. 

\item High resolution spectroscopy using multiple visible emission lines\,--\,[Fe x], [Fe xi], [Fe xiii] and [Fe xiv] need to be planned. There is no blending in these visible emission lines whereas crowding of the emission lines in EUV and UV range leads to blending. The corona is highly inhomogeneous in nature. The Figure \ref{fig:4} shows that intensity images of a coronal region in four emission lines are similar with some differences. The images in [Fe x] and [Fe xi] resemble more and images in [Fe iii] and [Fe xiv] are more correlated.  Since, the emission at the [Fe x] and [Fe xi] comes from the same plasma volume as these lines represent similar temperatures. Similarly, images in [Fe xiii] and [Fe xiv] resemble more as these lines represent similar coronal temperatures of 1.6\,${\times}$\,$10^{6}$\,K and 1.8\,${\times}$\,$10^{6}$\,K. Hence, the intensity ratios of [Fe xi] / [Fe x] and [Fe xiv] / [Fe xiii] will provide reliable values of temperature of the respective plasma, thereby that of coronal structures and their variations along the structures. Using these temperature values it will be possible to determine the reliable values of non-thermal velocity from the measured values of line width. This will help us to investigate the physical and dynamical characteristics coronal structures and understand the heating of coronal plasma.
Simultaneous multiple emission line spectroscopy with DKIST, if possible, would be useful in this connection.

\item Relatively more photons available in visible emission lines and possible high spectral resolution permit to derive accurate parameters of emission lines. Makes it possible to investigate their variation with height and in different coronal structures.

\item Measurement of coronal magnetic fields is very important. Several results have been published using ground based observations in infrared [Fe xiii] and radio wavelengths \citep{Kakinuma1962,Dulk_&_Mclean_1978,lin2000,Bastian2001,Vrsnak_2002,Gelfreikh2004,
Ryabov2004,White2004,Sastry_2009,Ramesh_2010_grip,ramesh2021,
Gopalswamy2012,hariharan2016,Alissandrakis2021,Yang2024,schad2024b,Sayuf2026b}. It is weak, and difficult to determine in infrared wavelengths due to varying scattered light on ground and in the instruments.  Dedicated efforts need to be made to measure magnetic fields from space in the [Fe xiii] emission line, with probably single instrument onboard and maximum use of available resources. 

\end{itemize}

\section{Conclusions}
 We have discussed the spectroscopy of the solar corona and its importance in detail. How the spectroscopy developed with time due to enhancement of available technology from photography to digital cameras. Extended  observations of solar corona from few minutes during the eclipses, for longer periods limited by sky conditions using ground based coronagraphs, and then continuous observations without the day-night cycle from space has played an important role to investigate the energetic events occurring on Sun and in corona. We have shown the need to perform the spectroscopy in multiple emission lines representing different coronal temperatures, especially in four bright emission lines in the visible part of the coronal spectrum using the high resolution instruments from space. The long duration observations in these emission lines will help us to understand the physical and dynamical characteristics of solar corona and  heating of coronal plasma. Spectroscopy in the UV, EUV and X-rays is also required irrespective of blending of most of the lines in this spectral band. It may be noted that in the visible part of the spectrum does not have the problem of blending of emission lines.

\section{Acknowledgements}

We thank the VELC technical and scientific  teams of Indian Institute of Astrophysics for their support to the project. We also thank the large number of staff at various ISRO organizations for their dedicated work to realize the instrument and launch it. Jagdev Singh thanks B. Raghavendra Prasad, Venkata Suresh Narra, P. U. Kamath and V. Natarajan for their dedication and hard work for the successful completion of the VELC project. We acknowledge the referees for their comments and suggestions.

\bibliography{reference} 

@ARTICLE{akiyama2005,
  author = {{Akiyama}, S. and {Doschek}, G.~A. and {Mariska}, J.~T.},
  title = "{Nonthermal Velocities in Different Temperature Regions of the Solar Lower Transition Region}",
  journal = {ApJ},
  volume = {623},
  pages = {540-550},
  year = {2005},
  doi = {10.1086/428638},
  adsurl = {https://ui.adsabs.harvard.edu/abs/2005ApJ...623..540A},
  adsbibcode = {2005ApJ...623..540A}
}

@ARTICLE{aly1962,
       author = {{Aly}, M.~K. and {Evans}, J.~W. and {Orrall}, F.~Q.},
        title = "{A Photometric Study of the Continuum and Seventeen Emission Lines in the Inner Solar Corona.}",
      journal = {ApJ},
         year = 1962,
        month = nov,
       volume = {136},
        pages = {956},
          doi = {10.1086/147449},
       adsurl = {https://ui.adsabs.harvard.edu/abs/1962ApJ...136..956A}
}

@ARTICLE{bappu1972,
       author = {{Bappu}, M.~K.~V. and {Bhattacharyya}, J.~C. and {Sivaraman}, K.~R.},
        title = "{On Emission Lines of Hydrogen, Helium and Ionized Calcium Seen on a Coronal Spectrogram of the March 7, 1970 Eclipse}",
      journal = {Sol. Phys.},
         year = 1972,
        month = oct,
       volume = {26},
       number = {2},
        pages = {366-369},
          doi = {10.1007/BF00165277},
       adsurl = {https://ui.adsabs.harvard.edu/abs/1972SoPh...26..366B}
}

@article{benz2009,
  title={A world-wide net of solar radio spectrometers: e-CALLISTO},
  author={Benz, AO and Monstein, C and Meyer, H and Manoharan, PK and Ramesh, R and Altyntsev, A and Lara, A and Paez, J and Cho, K-S},
  journal={Earth, Moon, \& Planets},
  volume={104},
  number={1},
  pages={277--285},
  year={2009},
  publisher={Springer},
  doi = {10.1007/s11038-008-9267-6},
  url = {https://doi.org/10.1007/s11038-008-9267-6}
}

@ARTICLE{brooks2016,
       author = {Brooks, David H. and Warren, Harry P.},
       title = "{Measurements of Non-thermal Line Widths in Solar Active Regions}",
       journal = {ApJ},
       year = {2016},
       volume = {820},
       number = {1},
       eid = {63},
       pages = {63},
       doi = {10.3847/0004-637X/820/1/63},
       archivePrefix = {arXiv},
       eprint = {1511.02313},
       primaryClass = {astro-ph.SR},
       adsurl = {https://ui.adsabs.harvard.edu/abs/2016ApJ...820...63B}
}

@ARTICLE{chandrasekhar1991,
       author = {{Chandrasekhar}, T. and {Desai}, J.~N. and {Ashok}, N.~M. and {Pasachoff}, Jay M.},
        title = "{Fabry-P{\'e}rot line profiles in the {\ensuremath{\lambda}}5303 {\r{a}} and {\ensuremath{\lambda}}6374 {\r{a}} coronal lines obtained during the 1983 Indonesian eclipse}",
      journal = {Sol. Phys.},
         year = 1991,
        month = jan,
       volume = {131},
       number = {1},
        pages = {25-39},
          doi = {10.1007/BF00151741},
       adsurl = {https://ui.adsabs.harvard.edu/abs/1991SoPh..131...25C}
}

@ARTICLE{contesse2004,
       author = {{Contesse}, L. and {Koutchmy}, S. and {Viladrich}, C.},
        title = "{Analysis of non-thermal velocities in the solar corona}",
      journal = {Ann. Geophys.},
         year = 2004,
        month = aug,
       volume = {22},
       number = {8},
        pages = {3055-3062},
          doi = {10.5194/angeo-22-3055-2004},
       adsurl = {https://ui.adsabs.harvard.edu/abs/2004AnGeo..22.3055C}
}

@ARTICLE{delzanna2018,
       author = {{Del Zanna}, Giulio and {Mason}, Helen E.},
        title = "{Solar UV and X-ray spectral diagnostics}",
      journal = {Liv. Rev. Sol. Phys.},
         year = 2018,
        month = dec,
       volume = {15},
       number = {1},
          eid = {5},
        pages = {5},
          doi = {10.1007/s41116-018-0015-3},
archivePrefix = {arXiv},
       eprint = {1809.01618},
 primaryClass = {astro-ph.SR},
       adsurl = {https://ui.adsabs.harvard.edu/abs/2018LRSP...15....5D}
}

@ARTICLE{desai1982,
       author = {{Desai}, J.~N. and {Chandrasekhar}, T. and {Angreji}, P.~D.},
        title = "{Doppler shift measurements on the green coronal line{\textemdash}evidence for largescale macroscopic mass motion}",
      journal = {J. Astrophys. Astr.},
         year = 1982,
        month = mar,
       volume = {3},
       number = {1},
        pages = {69-77},
          doi = {10.1007/BF02714564},
       adsurl = {https://ui.adsabs.harvard.edu/abs/1982JApA....3...69D}
}

@ARTICLE{dima2019,
       author = {{Dima}, Gabriel I. and {Kuhn}, Jeffrey R. and {Schad}, Thomas A.},
        title = "{Coronagraphic Observations of Si X {\ensuremath{\lambda}}14301 and Fe XIII {\ensuremath{\lambda}}10747 Linearly Polarized Spectra Using the SOLARC Telescope}",
      journal = {ApJ},
         year = 2019,
        month = jun,
       volume = {877},
       number = {2},
          eid = {144},
        pages = {144},
          doi = {10.3847/1538-4357/ab1cb7},
       adsurl = {https://ui.adsabs.harvard.edu/abs/2019ApJ...877..144D}
}

@ARTICLE{edlen1943,
       author = {{Edl{\'e}n}, Bengt},
        title = "{Die Deutung der Emissionslinien im Spektrum der Sonnenkorona. Mit 6 Abbildungen.}",
      journal = {Z. Astrophys.},
         year = 1943,
        month = jan,
       volume = {22},
        pages = {30},
       adsurl = {https://ui.adsabs.harvard.edu/abs/1943ZA.....22...30E}
}

@ARTICLE{giersch2017,
       author = {{Giersch}, O.~D. and {Kennewell}, J. and {Lynch}, M.},
        title = "{Solar Radio Burst Statistics and Implications for Space Weather Effects}",
      journal = {Space Weather},
         year = 2017,
        month = nov,
       volume = {15},
       number = {11},
        pages = {1511-1522},
          doi = {10.1002/2017SW001658},
       adsurl = {https://ui.adsabs.harvard.edu/abs/2017SpWea..15.1511G}
}

@INCOLLECTION{goldberg1948,
       author = {{Goldberg}, Leo and {Menzel}, Donald H.},
        title = "{The Solar Corona and Ultraviolet Radiation}",
    booktitle = {Contributions on Interstellar Matter, Electronic and Computational Devices, Eclipsing Binaries, The Gaseous Envelope of the Earth},
    publisher = {Harvard Observatory Monographs},
         year = 1948,
       volume = {7},
        pages = {279},
       adsurl = {https://ui.adsabs.harvard.edu/abs/1948HarMo...7..279G}
}

@ARTICLE{hahn2014,
       author = {{Hahn}, M. and {Savin}, D.~W.},
        title = "{Evidence for Wave Heating of the Quiet-Sun Corona}",
      journal = {ApJ},
         year = 2014,
        month = nov,
       volume = {795},
       number = {2},
          eid = {111},
        pages = {111},
          doi = {10.1088/0004-637X/795/2/111},
archivePrefix = {arXiv},
       eprint = {1407.3250},
 primaryClass = {astro-ph.SR},
       adsurl = {https://ui.adsabs.harvard.edu/abs/2014ApJ...795..111H}
}

@ARTICLE{harra1999,
       author = {{Harra-Murnion}, L.~K. and {Matthews}, S.~A. and {Hara}, H. and {Ichimoto}, K.},
        title = "{Dynamics of solar active region loops}",
      journal = {A\&A},
         year = 1999,
        month = may,
       volume = {345},
        pages = {1011-1018},
       adsurl = {https://ui.adsabs.harvard.edu/abs/1999A&A...345.1011H}
}

@ARTICLE{hara1999,
       author = {{Hara}, H. and {Ichimoto}, K.},
        title = "{Microscopic Nonthermal Plasma Motions of Coronal Loops in a Solar Active Region}",
      journal = {ApJ},
         year = 1999,
        month = mar,
       volume = {513},
       number = {2},
        pages = {969-982},
          doi = {10.1086/306880},
       adsurl = {https://ui.adsabs.harvard.edu/abs/1999ApJ...513..969H}
}

@article{harrison1995,
  title={The coronal diagnostic spectrometer for the solar and heliospheric observatory},
  author={Harrison, RA and Sawyer, EC and Carter, MK and Cruise, AM and Cutler, RM and Fludra, A and Hayes, RW and Kent, BJ and Lang, J and Parker, DJ and others},
  journal={Sol. Phys.},
  volume={162},
  number={1},
  pages={233--290},
  year={1995},
  publisher={Springer},
  url = {https://doi.org/10.1007/BF00733431},
  doi  = {10.1007/BF00733431}
}

@ARTICLE{ichimoto1995,
       author = {{Ichimoto}, K. and {Hara}, H. and {Takeda}, A. and {Kumagai}, K. and {Sakurai}, T. and {Shimizu}, T. and {Hudson}, H.~S.},
        title = "{Coordinated Observation of the Solar Corona Using the Norikura Coronagraph and the YOHKOH Soft X-Ray Telescope}",
      journal = {ApJ},
         year = 1995,
        month = jun,
       volume = {445},
        pages = {978},
          doi = {10.1086/175756},
       adsurl = {https://ui.adsabs.harvard.edu/abs/1995ApJ...445..978I}
}

@ARTICLE{ichimoto1996,
       author = {{Ichimoto}, Kiyoshi and {Kumagai}, Kazuyoshi and {Sano}, Issei and {Kobiki}, Toshihiko and {Munoz}, Alberto and {Sakurai}, Takashi},
        title = "{Measurement of the Coronal Electron Temperature at the Total Solar Eclipse on 1994 November 3}",
      journal = {PASJ},
         year = 1996,
        month = jun,
       volume = {48},
       number = {3},
        pages = {545-554},
          doi = {10.1093/pasj/48.3.545},
       adsurl = {https://ui.adsabs.harvard.edu/abs/1996PASJ...48..545I}
}

@ARTICLE{iwai2017,
       author = {{Iwai}, Kazumasa and {Kubo}, Y{\^u}ki and {Ishibashi}, Hiromitsu and {Naoi}, Takahiro and {Harada}, Kenichi and {Ema}, Kenji and {Hayashi}, Yoshinori and {Chikahiro}, Yuichi},
        title = "{OCTAD-S: digital fast Fourier transform spectrometers by FPGA}",
      journal = {Earth, Planets, \& Space},
         year = 2017,
        month = dec,
       volume = {69},
       number = {1},
          eid = {95},
        pages = {95},
          doi = {10.1186/s40623-017-0681-8},
archivePrefix = {arXiv},
       eprint = {1706.08863},
 primaryClass = {astro-ph.IM},
       adsurl = {https://ui.adsabs.harvard.edu/abs/2017EP&S...69...95I}
}

@ARTICLE{jarrett1955,
       author = {{Jarrett}, A.~H. and {von Kl{\"u}ber}, H.},
        title = "{Interferometric measurements of the green corona line during the total solar eclipse of 1954 June 30}",
      journal = {MNRAS},
         year = 1955,
        month = jan,
       volume = {115},
        pages = {343},
          doi = {10.1093/mnras/115.4.343},
       adsurl = {https://ui.adsabs.harvard.edu/abs/1955MNRAS.115..343J}
}

@ARTICLE{judge2002,
       author = {{Judge}, P.~G. and {Tomczyk}, S. and {Livingston}, W.~C. and {Keller}, C.~U. and {Penn}, M.~J.},
        title = "{Spectroscopic Detection of the 3.934 Micron Line of Si IX in the Solar Corona}",
      journal = {ApJL},
         year = 2002,
        month = sep,
       volume = {576},
       number = {2},
        pages = {L157-L160},
          doi = {10.1086/343730},
       adsurl = {https://ui.adsabs.harvard.edu/abs/2002ApJ...576L.157J}
}

@ARTICLE{koutchmy1983,
       author = {{Koutchmy}, S. and {Zhugzhda}, Ia. D. and {Locans}, V.},
        title = "{Short period coronal oscillations - Observation and interpretation}",
      journal = {A\&A},
         year = 1983,
        month = apr,
       volume = {120},
       number = {2},
        pages = {185-191},
       adsurl = {https://ui.adsabs.harvard.edu/abs/1983A&A...120..185K}
}

@ARTICLE{kathiravan2026,
       author = {{Kathiravan}, C. and {Muthu Priyal}, V. and {Ramesh}, R.},
        title = "{Constraining the onset height of coronal mass ejection driven shocks using near-Sun observations in visible and radio wavelengths}",
      journal = {J. Astrophys. Astr.},
         year = 2026,
        month = apr,
       volume = {47},
       number = {1},
          eid = {29},
        pages = {29},
          doi = {10.1007/s12036-026-10147-2},
       adsurl = {https://ui.adsabs.harvard.edu/abs/2026JApA...47...29K}
}

@ARTICLE{koutchmy2019,
       author = {{Koutchmy}, S. and {Baudin}, F. and {Abdi}, Sh. and {Golub}, L. and {S{\`e}vre}, F.},
        title = "{New deep coronal spectra from the 2017 total solar eclipse}",
      journal = {A\&A},
         year = 2019,
        month = dec,
       volume = {632},
          eid = {A86},
        pages = {A86},
          doi = {10.1051/0004-6361/201935681},
archivePrefix = {arXiv},
       eprint = {1910.01372},
 primaryClass = {astro-ph.SR},
       adsurl = {https://ui.adsabs.harvard.edu/abs/2019A&A...632A..86K}
}

@ARTICLE{kuhn1996,
       author = {{Kuhn}, J.~R. and {Penn}, M.~J. and {Mann}, I.},
        title = "{The Near-Infrared Coronal Spectrum}",
      journal = {ApJL},
         year = 1996,
        month = jan,
       volume = {456},
        pages = {L67},
          doi = {10.1086/309864},
       adsurl = {https://ui.adsabs.harvard.edu/abs/1996ApJ...456L..67K}
}

@INPROCEEDINGS{liebenberg1974,
       author = {{Liebenberg}, D.~H. and {Hoffman}, M.~M.},
        title = "{Temporal Observations of the {\ensuremath{\lambda}}'' 5303 Emission Line Profile during the 74 Minute Totality from the Concorde SST at the 30 June 1973 Total Solar Eclipse: Preliminary Intensity Variations above an Active Region (presented by D. H. Liebenberg)}",
    booktitle = {Coronal Disturbances},
         year = 1974,
       editor = {{Newkirk}, Gordon Allen},
       series = {IAU Symposium},
       volume = {57},
        month = jan,
        pages = {485},
       adsurl = {https://ui.adsabs.harvard.edu/abs/1974IAUS...57..485L}
}

@ARTICLE{lin2004,
       author = {{Lin}, H. and {Kuhn}, J.~R. and {Coulter}, R.},
        title = "{Coronal Magnetic Field Measurements}",
      journal = {ApJL},
         year = 2004,
        month = oct,
       volume = {613},
       number = {2},
        pages = {L177-L180},
          doi = {10.1086/425217},
       adsurl = {https://ui.adsabs.harvard.edu/abs/2004ApJ...613L.177L}
}

@article{livingston1970,
  author={Livingston, W and  Harvey, J and Doe, L.},
  pages={72},
  year={1970},
  journal={Solar Eclipse 1970 Bulletin F, National Science Foundation, Washington}
}

@ARTICLE{lyot1955,
       author = {{Lyot}, B. and {Aly}, M.~K.},
        title = "{Preliminary Note on Measures of Coronal Emission Lines Observed at Total Solar Eclipse, February 25, 1952, by}",
      journal = {ApJ},
         year = 1955,
        month = nov,
       volume = {122},
        pages = {438},
          doi = {10.1086/146105},
       adsurl = {https://ui.adsabs.harvard.edu/abs/1955ApJ...122..438L}
}

@ARTICLE{mierla2008,
       author = {{Mierla}, M. and {Schwenn}, R. and {Teriaca}, L. and {Stenborg}, G. and {Podlipnik}, B.},
        title = "{Analysis of the Fe X and Fe XIV line width in the solar corona using LASCO-C1 spectral data}",
      journal = {A\&A},
         year = 2008,
        month = mar,
       volume = {480},
       number = {2},
        pages = {509-514},
          doi = {10.1051/0004-6361:20078329},
archivePrefix = {arXiv},
       eprint = {0903.0496},
 primaryClass = {astro-ph.SR},
       adsurl = {https://ui.adsabs.harvard.edu/abs/2008A&A...480..509M}
}

@article{parate2025,
  title={Aditya-L1-An Observatory Class Mission: Spacecraft, Mission, and Operations},
  author={Parate, Nashiket Premlal and Shaji, Nigar and Srikanth, Motamarri and Sankarasubramanian, K and Patnaik, Narashimha Murthy and Narendra, S and Adoni, Abhijit Avinash and Negi, Kuldeep and Pattanaik, Pragyan and GV P, Bharat Kumar and others},
  journal={Sol. Phys.},
  volume={300},
  number={9},
  pages={128},
  year={2025},
  publisher={Springer},
  doi = {10.1007/s11207-025-02533-w},
  url = {https://doi.org/10.1007/s11207-025-02533-w}
}

@ARTICLE{prasad2013,
       author = {{Krishna Prasad}, S. and {Singh}, Jagdev and {Ichimoto}, K.},
        title = "{Thermal Structure of Coronal Loops as Seen with Norikura Coronagraph}",
      journal = {ApJL},
         year = 2013,
        month = mar,
       volume = {765},
       number = {2},
          eid = {L46},
        pages = {L46},
          doi = {10.1088/2041-8205/765/2/L46},
archivePrefix = {arXiv},
       eprint = {1302.5905},
 primaryClass = {astro-ph.SR},
       adsurl = {https://ui.adsabs.harvard.edu/abs/2013ApJ...765L..46K}
}

@ARTICLE{priyal2025a,
       author = {{Muthu Priyal}, V. and {Ramesh}, R. and {Singh}, Jagdev and {Sasikumar Raja}, K.},
        title = "{Observations of Temporal Variations in the 5303 {\r{A}} Emission Line During a Coronal Mass Ejection Without Any Associated Solar Flare}",
      journal = {ApJ},
         year = 2025,
        month = apr,
       volume = {983},
       number = {2},
          eid = {171},
        pages = {171},
          doi = {10.3847/1538-4357/adbe73},
       adsurl = {https://ui.adsabs.harvard.edu/abs/2025ApJ...983..171M}
}

@ARTICLE{priyal2025b,
       author = {{Muthu Priyal}, V. and {Ramesh}, R. and {Singh}, Jagdev and {Sasikumar Raja}, K. and {Gopalswamy}, N.},
        title = "{Near-Sun Characteristics of a CME Inferred from Observations in the 5303 {\r{A}} Coronal Emission Line}",
      journal = {ApJ},
         year = 2025,
        month = dec,
       volume = {994},
       number = {2},
          eid = {182},
        pages = {182},
          doi = {10.3847/1538-4357/ae1a47},
       adsurl = {https://ui.adsabs.harvard.edu/abs/2025ApJ...994..182P}
}

@article{raju1993,
    author = {Raju, K. P. and Desai, J. N. and Chandrasekhar, T. and Ashok, N. M.},
    title = {Line-of-sight velocities observed in the inner solar corona during the total solar eclipses of 1980 and 1983},
    journal = {MNRAS},
    volume = {263},
    number = {3},
    pages = {789-797},
    year = {1993},
    month = {08},
    issn = {0035-8711},
    doi = {10.1093/mnras/263.3.789},
    url = {https://doi.org/10.1093/mnras/263.3.789},
    eprint = {https://academic.oup.com/mnras/article-pdf/263/3/789/4040000/mnras263-0789.pdf},
}

@article{raju2000,
doi = {10.1086/317143},
url = {https://doi.org/10.1086/317143},
year = {2000},
month = {nov},
publisher = {},
volume = {543},
number = {2},
pages = {1044},
author = {Raju, K. P. and Sakurai, T. and Ichimoto, K. and Singh, Jagdev},
title = {The Physical Conditions in a Polar Coronal Hole and
Nearby Regions from Norikura
and SOHO Observations},
journal = {ApJ}
}

@INPROCEEDINGS{ramesh2011,
   author = {{Ramesh}, R.},
    title = "{Low frequency solar radio astronomy at the Indian Institute of Astrophysics}",
    booktitle = {1st Asia-Pacific Sol. Phys. Meeting},
     series = {Astron. Soc. India Conf. Ser. 2},
    editor = {{Choudhuri}, A. R. and {Banerjee}, D.},
    year = 2011,
     month = jan,
     pages = {55},
     adsurl = {http://cdsads.u-strasbg.fr/abs/2011ASInC...2...55R}
}

@ARTICLE{ramesh2012,
   author = {{Ramesh}, R. and {Kathiravan}, C. and {Indrajit Barve}, V. and {Rajalingam}, M.},
    title = "{High Angular Resolution Radio Observations of a Coronal Mass Ejection Source Region at Low Frequencies during a Solar Eclipse}",
  journal = {ApJ},
     year = {2012},
    month = jan,
   volume = 744,
    pages = {165},
      doi = {10.1088/0004-637X/744/2/165},
   adsurl = {http://cdsads.u-strasbg.fr/abs/2012ApJ...744..165R}
}

@ARTICLE{ramesh2024,
       author = {{Ramesh}, R. and {Muthu Priyal}, V. and {Singh}, Jagdev and {Sasikumar Raja}, K. and {Savarimuthu}, P. and {Gavshinde}, Priya},
        title = "{New Results on the Onset of a Coronal Mass Ejection from 5303 {\r{A}} Emission Line Observations with VELC/ADITYA-L1}",
      journal = {ApJL},
         year = 2024,
        month = nov,
       volume = {976},
       number = {1},
          eid = {L6},
        pages = {L6},
          doi = {10.3847/2041-8213/ad8c45},
archivePrefix = {arXiv},
       eprint = {2411.09408},
 primaryClass = {astro-ph.SR},
       adsurl = {https://ui.adsabs.harvard.edu/abs/2024ApJ...976L...6R}
}

@ARTICLE{Ramesh2026a,
   author = {{Ramesh}, R. and {Muthu Priyal}, V. and {Singh}, J. and {Sasikumar Raja}, K. and {Mishra}, W. and {Monstein}, C.},
    title = "{Estimation of the Energy Involved in a Solar Coronal Mass Ejection Using Spectroscopic Observations in the 5303{\AA} Emission Line}",
    journal = {ApJL},
     year = {2026},
     month = jul,
    volume = 1005,
    pages = {L7},
    doi = {10.3847/2041-8213/ae7bf0},
     adsurl = {https://ui.adsabs.harvard.edu/abs/2026ApJ..1005L...7R/abstract}
}

@ARTICLE{Ramesh2026b,
   author = {{Ramesh}, R. and {Muthu Priyal}, V. and {Singh}, J. and {Sasikumar Raja}, K. and {Savarimuthu}, P. and {Yuvashree}, E. and {Nagashree}, S.},
    title = "{Spectroscopic Observations of 5303{\AA} Emission along a Preflare Solar Coronal Loop - Evidence for Plasma Upflows}",
    journal = {ApJ},
     year = {2026},
     month = aug,
    volume = 1007,
    pages = {144},
    doi = {10.3847/1538-4357/ae8ba7},
     adsurl = {https://ui.adsabs.harvard.edu/abs/2026ApJ..1007..144R/abstract}
}

@ARTICLE{rimmele2020,
       author = {{Rimmele}, Thomas R. and {Warner}, Mark and {Keil}, Stephen L. and {Goode}, Philip R. and {Kn{\"o}lker}, Michael and {Kuhn}, Jeffrey R. and {Rosner}, Robert R. and {McMullin}, Joseph P. and {Casini}, Roberto and {Lin}, Haosheng and {W{\"o}ger}, Friedrich and {von der L{\"u}he}, Oskar and {Tritschler}, Alexandra and {Davey}, Alisdair and {de Wijn}, Alfred and {Elmore}, David F. and {Fehlmann}, Andr{\'e} and {Harrington}, David M. and {Jaeggli}, Sarah A. and {Rast}, Mark P. and {Schad}, Thomas A. and {Schmidt}, Wolfgang and {Mathioudakis}, Mihalis and {Mickey}, Donald L. and {Anan}, Tetsu and {Beck}, Christian and {Marshall}, Heather K. and {Jeffers}, Paul F. and {Oschmann}, Jacobus M. and {Beard}, Andrew and {Berst}, David C. and {Cowan}, Bruce A. and {Craig}, Simon C. and {Cross}, Eric and {Cummings}, Bryan K. and {Donnelly}, Colleen and {de Vanssay}, Jean-Benoit and {Eigenbrot}, Arthur D. and {Ferayorni}, Andrew and {Foster}, Christopher and {Galapon}, Chriselle Ann and {Gedrites}, Christopher and {Gonzales}, Kerry and {Goodrich}, Bret D. and {Gregory}, Brian S. and {Guzman}, Stephanie S. and {Guzzo}, Stephen and {Hegwer}, Steve and {Hubbard}, Robert P. and {Hubbard}, John R. and {Johansson}, Erik M. and {Johnson}, Luke C. and {Liang}, Chen and {Liang}, Mary and {McQuillen}, Isaac and {Mayer}, Christopher and {Newman}, Karl and {Onodera}, Brialyn and {Phelps}, LeEllen and {Puentes}, Myles M. and {Richards}, Christopher and {Rimmele}, Lukas M. and {Sekulic}, Predrag and {Shimko}, Stephan R. and {Simison}, Brett E. and {Smith}, Brett and {Starman}, Erik and {Sueoka}, Stacey R. and {Summers}, Richard T. and {Szabo}, Aimee and {Szabo}, Louis and {Wampler}, Stephen B. and {Williams}, Timothy R. and {White}, Charles},
        title = "{The Daniel K. Inouye Solar Telescope - Observatory Overview}",
      journal = {Sol. Phys.},
         year = 2020,
        month = dec,
       volume = {295},
       number = {12},
          eid = {172},
        pages = {172},
          doi = {10.1007/s11207-020-01736-7},
       adsurl = {https://ui.adsabs.harvard.edu/abs/2020SoPh..295..172R}
}

@ARTICLE{sakurai2002,
       author = {{Sakurai}, T. and {Ichimoto}, K. and {Raju}, K.~P. and {Singh}, J.},
        title = "{Spectroscopic Observation of Coronal Waves}",
      journal = {Sol. Phys.},
         year = 2002,
        month = oct,
       volume = {209},
       number = {2},
        pages = {265-286},
          doi = {10.1023/A:1021297313448},
       adsurl = {https://ui.adsabs.harvard.edu/abs/2002SoPh..209..265S}
}

@ARTICLE{schad2024a,
       author = {{Schad}, Thomas A. and {Fehlmann}, Andre and {Dima}, Gabriel I. and {Kuhn}, Jeffrey R. and {Scholl}, Isabelle F. and {Harrington}, David and {Rimmele}, Thomas and {Tritschler}, Alexandra and {Paraschiv}, Alin R.},
        title = "{Coronagraphic Observations of Si X 1430 nm Acquired by DKIST/Cryo-NIRSP with Methods for Telluric Absorption Correction}",
      journal = {ApJ},
         year = 2024,
        month = apr,
       volume = {965},
       number = {1},
          eid = {40},
        pages = {40},
          doi = {10.3847/1538-4357/ad2995},
archivePrefix = {arXiv},
       eprint = {2402.09632},
 primaryClass = {astro-ph.SR},
       adsurl = {https://ui.adsabs.harvard.edu/abs/2024ApJ...965...40S}
}

@ARTICLE{schad2024b,
       author = {{Schad}, Thomas A. and {Petrie}, Gordon and {Kuhn}, Jeffrey and {Fehlmann}, Andre and {Rimmele}, Thomas and {Tritschler}, Alexandra and {Woeger}, Friedrich and {Scholl}, Isabelle and {Williams}, Rebecca and {Harrington}, David and {Paraschiv}, Alin and {Szente}, Judit},
        title = "{Mapping the Sun's coronal magnetic field using the Zeeman effect}",
      journal = {Sci. Adv.},
         year = 2024,
        month = sep,
       volume = {10},
       number = {37},
        pages = {eadq1604},
          doi = {10.1126/sciadv.adq1604},
archivePrefix = {arXiv},
       eprint = {2410.21568},
 primaryClass = {astro-ph.SR},
       adsurl = {https://ui.adsabs.harvard.edu/abs/2024SciA...10.1604S}
}

@article{singh1982,
  title={Eclipse observations of coronal emission lines. I.[Fex] 6374{\AA} profiles at the eclipse of 16 February 1980},
  author={Singh, Jagdev and Bappu, MKV and Saxena, AK},
  journal={J. Astrophys. Astr.},
  volume={3},
  number={3},
  pages={249--266},
  year={1982},
  publisher={Springer},
  doi = {10.1007/BF02714865},
  url = {https://doi.org/10.1007/BF02714865}
}

@ARTICLE{singh1985,
       author = {{Singh}, J.},
        title = "{Clues to the mode of excitation of Fe x ions in the solar corona from the 1980 eclipse observations}",
      journal = {Sol. Phys.},
         year = 1985,
        month = feb,
       volume = {95},
       number = {2},
        pages = {253-262},
          doi = {10.1007/BF00152403},
       adsurl = {https://ui.adsabs.harvard.edu/abs/1985SoPh...95..253S}
}

@ARTICLE{singh1999,
       author = {{Singh}, Jagdev and {Ichimoto}, Kiyoshi and {Imai}, Hideki and {Sakurai}, Takashi and {Takeda}, Aki},
        title = "{Spectroscopic Studies of the Solar Corona I. Spatial Variations in Line Parameters of Green and Red Coronal Lines}",
      journal = {PASJ},
         year = 1999,
        month = apr,
       volume = {51},
       number = {2},
        pages = {269-276},
          doi = {10.1093/pasj/51.2.269},
       adsurl = {https://ui.adsabs.harvard.edu/abs/1999PASJ...51..269S}
}

@ARTICLE{singh2002,
       author = {{Singh}, Jagdev and {Sakurai}, Takashi and {Ichimoto}, Kiyoshi and {Suematsu}, Yoshinori and {Takeda}, Aki},
        title = "{Spectroscopic Studies of the Solar Corona II. Properties of Green and Red Emission Lines in Open and Closed Coronal Structures}",
      journal = {PASJ},
         year = 2002,
        month = oct,
       volume = {54},
        pages = {793-806},
          doi = {10.1093/pasj/54.5.793},
       adsurl = {https://ui.adsabs.harvard.edu/abs/2002PASJ...54..793S}
}

@ARTICLE{singh2003a,
       author = {{Singh}, Jagdev and {Ichimoto}, Kiyoshi and {Sakurai}, Takashi and {Muneer}, S.},
        title = "{Spectroscopic Studies of the Solar Corona. IV. Physical Properties of Coronal Structure}",
      journal = {ApJ},
         year = 2003,
        month = mar,
       volume = {585},
       number = {1},
        pages = {516-523},
          doi = {10.1086/346000},
       adsurl = {https://ui.adsabs.harvard.edu/abs/2003ApJ...585..516S}
}

@ARTICLE{singh2003b,
       author = {{Singh}, Jagdev and {Sakurai}, Takashi and {Ichimoto}, Kiyoshi and {Muneer}, S.},
        title = "{Spectroscopic Studies of the Solar Corona - V. Physical Properties of Coronal Structures}",
      journal = {Sol. Phys.},
         year = 2003,
        month = feb,
       volume = {212},
       number = {2},
        pages = {343-359},
          doi = {10.1023/A:1022948520902},
       adsurl = {https://ui.adsabs.harvard.edu/abs/2003SoPh..212..343S}
}

@ARTICLE{singh2004,
       author = {{Singh}, Jagdev and {Sakurai}, Takashi and {Ichimoto}, Kiyoshi and {Watanabe}, Tetsuya},
        title = "{Complex Variations in the Line-Intensity Ratio of Coronal Emission Lines with Height above the Limb}",
      journal = {ApJL},
         year = 2004,
        month = dec,
       volume = {617},
       number = {1},
        pages = {L81-L84},
          doi = {10.1086/427078},
       adsurl = {https://ui.adsabs.harvard.edu/abs/2004ApJ...617L..81S}
}

@ARTICLE{singh2005,
       author = {{Singh}, Jagdev and {Sakurai}, Takashi and {Ichimoto}, Kiyoshi and {Suzuki}, Isao and {Hagino}, Masaoki},
        title = "{Spectroscopic Studies of Solar Corona VII. Formation of a Coronal Loop by Evaporation}",
      journal = {Sol. Phys.},
         year = 2005,
        month = feb,
       volume = {226},
       number = {2},
        pages = {201-221},
          doi = {10.1007/s11207-005-7415-y},
       adsurl = {https://ui.adsabs.harvard.edu/abs/2005SoPh..226..201S}
}

@article{singh2006,
doi = {10.1086/499200},
url = {https://doi.org/10.1086/499200},
year = {2006},
month = {mar},
publisher = {},
volume = {639},
number = {1},
pages = {475},
author = {Singh, Jagdev and Sakurai, Takashi and Ichimoto, Kiyoshi},
title = {Do the Line Widths of Coronal Emission Lines Increase with Height above the Limb?},
journal = {ApJ},
}

@ARTICLE{singh2025,
       author = {{Singh}, Jagdev and {Ramesh}, R. and {Prasad}, B. Raghavendra and {Priyal}, V. Muthu and {Raja}, K. Sasikumar and {Venkata}, S.~N. and {Kamath}, P.~U. and {Natarajan}, V. and {Pawankumar}, S. and {Krishnan}, V.~U. Sanal and {Savarimuthu}, P. and {Mishra}, Shalabh and {Kumar}, Varun and {Sumana}, Chavali and {Hegde}, S. Bhavana and {Utkarsha}, D. and {Kumar}, Amit and {Nagabhushana}, S. and {Kathiravan}, S. and {Vemareddy}, P. and {Kathiravan}, C. and {Nagaraju}, K. and {Ravindra}, Belur and {Mishra}, Wageesh},
        title = "{Visible Emission Line Coronagraph (VELC) on Board Aditya-L1}",
      journal = {Sol. Phys.},
         year = 2025,
        month = may,
       volume = {300},
       number = {5},
          eid = {66},
        pages = {66},
          doi = {10.1007/s11207-025-02477-1},
archivePrefix = {arXiv},
       eprint = {2503.21184},
 primaryClass = {astro-ph.SR},
       adsurl = {https://ui.adsabs.harvard.edu/abs/2025SoPh..300...66S}
}

@ARTICLE{srivastava2025,
       author = {{Srivastava}, A.~K. and {Mondal}, Sripan and {Priest}, Eric R. and {Mishra}, Sudheer K. and {Pontin}, David I. and {Kwon}, R.~Y. and {Yuan}, Ding and {Murawski}, K. and {Asai}, Ayumi},
        title = "{Localized Heating and Dynamics of the Solar Corona due to a Symbiosis of Waves and Reconnection}",
      journal = {ApJ},
         year = 2025,
        month = may,
       volume = {984},
       number = {1},
          eid = {36},
        pages = {36},
          doi = {10.3847/1538-4357/adc379},
archivePrefix = {arXiv},
       eprint = {2503.16300},
 primaryClass = {astro-ph.SR},
       adsurl = {https://ui.adsabs.harvard.edu/abs/2025ApJ...984...36S}
}

@ARTICLE{sudheer2023,
       author = {{Mishra}, S. K. and {Sangal}, K. and {Kashyap}, P. and {Jel{\'\i}nek}, P. and {Srivastava}, A. K. and {Rajaguru}, S. P.},
        title = "{Origin of Quasi-periodic Pulsation at the Base of a Kink-unstable Jet}",
      journal = {ApJ},
         year = 2023,
        month = mar,
       volume = {945},
       number = {2},
          eid = {113},
        pages = {113},
          doi = {10.3847/1538-4357/acb058},
archivePrefix = {arXiv},
       eprint = {2301.01534},
 primaryClass = {astro-ph.SR},
       adsurl = {https://ui.adsabs.harvard.edu/abs/2023ApJ...945..113M}
}

@ARTICLE{tsubaki1975,
       author = {{Tsubaki}, T.},
        title = "{Line Profile Analysis of a Coronal Formation Observed near a Quiescent Prominence: Intensities, Temperatures and Velocity Fields}",
      journal = {Sol. Phys.},
         year = 1975,
        month = jul,
       volume = {43},
       number = {1},
        pages = {147-175},
          doi = {10.1007/BF00155151},
       adsurl = {https://ui.adsabs.harvard.edu/abs/1975SoPh...43..147T}
}

@ARTICLE{tsubaki1977,
       author = {{Tsubaki}, T.},
        title = "{Periodic oscillations found in coronal velocity fields.}",
      journal = {Sol. Phys.},
         year = 1977,
        month = jan,
       volume = {51},
       number = {1},
        pages = {121-130},
          doi = {10.1007/BF00240450},
       adsurl = {https://ui.adsabs.harvard.edu/abs/1977SoPh...51..121T}
}

@article{voulgaris2010,
  title={A comparison of the red and green coronal line intensities at the 29 March 2006 and the 1 August 2008 total solar eclipses: Considerations of the temperature of the solar corona},
  author={Voulgaris, Aristeidis and Athanasiadis, T and Seiradakis, JH and Pasachoff, JM},
  journal={Sol. Phys.},
  volume={264},
  number={1},
  pages={45--55},
  year={2010},
  publisher={Springer},
  url = {https://doi.org/10.1007/s11207-010-9575-7},
  doi  = {10.1007/s11207-010-9575-7}
}

@article{voulgaris2012,
  title={Spectroscopic coronal observations during the total solar eclipse of 11 July 2010},
  author={Voulgaris, Aristeidis G and Gaintatzis, PS and Seiradakis, JH and Pasachoff, JM and Economou, TE},
  journal={Sol. Phys.},
  volume={278},
  number={1},
  pages={187--202},
  year={2012},
  publisher={Springer},
  url  = {https://doi.org/10.1007/s11207-012-9929-4},
  doi  = {10.1007/s11207-012-9929-4}
}

@article{watanabe2007,
    author = {Watanabe, Tetsuya and Hara, Hirohisa and Culhane, Len and Harra, Louise K. and Doschek, George A. and Mariska, John T. and Young, Peter R.},
    title = {Temperature and Density Structures of Solar Corona—A Test of Iron Line Diagnostic Capability of EIS Instrument on Board Hinode},
    journal = {PASJ},
    volume = {59},
    number = {sp3},
    pages = {S669-S674},
    year = {2007},
    month = {11},
    doi = {10.1093/pasj/59.sp3.S669},
    url = {https://doi.org/10.1093/pasj/59.sp3.S669},
    eprint = {https://academic.oup.com/pasj/article-pdf/59/sp3/S669/17451762/pasj59-S669.pdf},
}

@article{ramesh1998,
  title={The gauribidanur radioheliograph},
  author={{Ramesh}, R. and {Subramanian}, K. R. and {Sundara Rajan}, M. S. and {Sastry}, {\relax Ch} V.},
  journal={Sol. Phys.},
  volume={181},
  number={2},
  pages={439--453},
  year={1998},
  publisher={Springer},
  url  = {https://doi.org/10.1023/A:1005075003370},
  doi  = {10.1023/A:1005075003370}
}

@INPROCEEDINGS{kerdraon1997,
   author = {{Kerdraon}, A. and {Delouis}, J. M.},
    title = "{The Nançay Radioheliograph}",
booktitle = {Coronal Physics from Radio and Space Observations},
     year = 1997,
   series = {Lecture Notes in Physics, Vol.483},
   editor = {{Trottet}, G.},
    pages = {192},
      doi = {10.1007/BFb0106458},
   adsurl = {https://ui.adsabs.harvard.edu/abs/1997LNP...483..192K/abstract}
}

@ARTICLE{hariharan2014,
   author = {{Hariharan}, K. and {Ramesh}, R. and {Kishore}, P. and {Kathiravan}, C. and 
	{Gopalswamy}, N.},
    title = "{An Estimate of the Coronal Magnetic Field near a Solar Coronal Mass Ejection from Low-frequency Radio Observations}",
  journal = {ApJ},
     year = 2014,
    month = nov,
   volume = 795,
      eid = {14},
    pages = {14},
      doi = {10.1088/0004-637X/795/1/14},
   adsurl = {http://adsabs.harvard.edu/abs/2014ApJ...795...14H}
}

@ARTICLE{ramesh2001,
   author = {{Ramesh}, R. and {Ebenezer}, E.},
    title = "{Decameter Wavelength Observations of an Absorption Burst from the Sun and Its Association with an X2.0/3B Flare and the Onset of a ``Halo'' Coronal Mass Ejection}",
  journal = {ApJL},
     year = {2001},
    month = sep,
   volume = 558,
    pages = {L141},
    doi = {10.1086/323498},
    adsurl = {https://ui.adsabs.harvard.edu/abs/2001ApJ...558L.141R/abstract}
}

@ARTICLE{kano1995,
   author = {{Kano}, R. and {Tsuneta}, S.},
    title = "{Scaling Law of Solar Coronal Loops Obtained with YOHKOH}",
  journal = {ApJ},
     year = {1995},
    month = dec,
   volume = 454,
    pages = {934},
    doi = {10.1086/176547},
    adsurl = {https://ui.adsabs.harvard.edu/abs/1995ApJ...454..934K/abstract}
}

@ARTICLE{yan2021,
   author = {{Yan}, Y. and {Chen}, Z. and {Wang}, W. and {Liu}, F. and {Geng}, L. and {Chen}, L. and
{Tan}, C. and {Chen}, X. and {Su}, C. and {Tan}, B.},
    title = "{Mingantu Spectral Radioheliograph for Solar and Space Weather Studies}",
  journal = {Front. Astron. Space Sci.},
     year = {2021},
   volume = 8,
    pages = {20},
    doi = {10.3389/fspas.2021.584043},
    adsurl = {https://ui.adsabs.harvard.edu/abs/2021FrASS...8...20Y/abstract}
}

@ARTICLE{lin2000,
   author = {{Lin}, H. and {Penn}, M.~J. and {Tomczyk}, S.},
    title = "{A New Precise Measurement of the Coronal Magnetic Field Strength}",
  journal = {ApJL},
     year = 2000,
    month = oct,
   volume = 541,
    pages = {L83-L86},
      doi = {10.1086/312900},
   adsurl = {http://adsabs.harvard.edu/abs/2000ApJ...541L..83L}
}

@ARTICLE{ramesh2021,
   author = {{Ramesh}, R. and {Kumari}, A. and {Kathiravan}, C. and {Ketaki}, D. and {Wang}, T. J.},
    title = "{New Results on the Direct Observations of Thermal Radio Emission from a Solar Coronal Mass Ejection}",
  journal = {Geophys. Res. Lett.},
     year = 2021,
   volume = 48,
    pages = {e91048},
    doi = {10.1029/2020GL091048},
   adsurl = {https://ui.adsabs.harvard.edu/abs/2021GeoRL..4891048R/abstract}
}

@ARTICLE{hariharan2016,
   author = {{Hariharan}, K. and {Ramesh}, R. and {Kathiravan}, C. and {Wang}, T. J.},
    title = "{Simultaneous Near-Sun Observations of a Moving Type IV Radio Burst and the Associated White-Light Coronal Mass Ejection}",
  journal = {Sol. Phys.},
     year = 2016,
   volume = 291,
    pages = {1405},
    doi = {10.1007/s11207-016-0918-x},
   adsurl = {https://ui.adsabs.harvard.edu/abs/2016SoPh..291.1405H/abstract}
}

@ARTICLE{Ramesh1999,
   author = {{Ramesh}, R. and {Subramanian}, K.~R. and {Sastry}, {\relax Ch} V.},
    title = "{Eclipse Observations of Compact Sources in the Outer Solar Corona}",
  journal = {Sol. Phys.},
     year = {1999},
    month = mar,
   volume = 185,
    pages = {77},
      doi = {10.1023/A:1005149830652},
   adsurl = {http://cdsads.u-strasbg.fr/abs/1999SoPh..185...77R}
}

@ARTICLE{Habbal2011,
   author = {{Habbal}, S. R. and {Druckm{\" u}ller}, M. and {Morgan}, H. and {Ding}, A. and {Johnson}, J. and {Druckm{\" u}llerov{\'a}}, H. and {Daw}, A. and {Arndt}, M. B. and
   {Dietzel}, M. and {Saken}, J.},
    title = "{Thermodynamics of the Solar Corona and Evolution of the Solar Magnetic Field as Inferred from the Total Solar Eclipse Observations of 2010 July 11}",
    journal = {ApJ},
     year = {2011},
     month = jun,
    volume = 734,
    pages = {120},    
    doi = {10.1088/0004-637X/734/2/120},
     adsurl = {https://ui.adsabs.harvard.edu/abs/2011ApJ...734..120H/abstract}
}

@ARTICLE{Boe2020,
   author = {{Boe}, B. and {Habbal}, S. and {Druckm{\" u}ller}, M. and {Ding}, A. and {Hod{\' e}rov}, J. and {{\v S}tarha}, P.},
    title = "{CME-induced Thermodynamic Changes in the Corona as Inferred from Fe XI and Fe XIV Emission Observations during the 2017 August 21 Total Solar Eclipse}",
    journal = {ApJ},
     year = {2020},
     month = jan,
    volume = 888,
    pages = {100},    
    doi = {10.3847/1538-4357/ab5e34},
     adsurl = {https://ui.adsabs.harvard.edu/abs/2020ApJ...888..100B/abstract}
}

@ARTICLE{Zhu2024,
   author = {{Zhu}, Y. and {Habbal}, S. R. and {Ding}, A. and {Yamashiro}, B. and {Landi}, E. and {Boe}, B. and {Constantinou}, S. and {Nassir}, M.},
    title = "{Spectroscopic Observations of the Solar Corona during the 2017 August 21 Total Solar Eclipse: Comparison of Spectral Line Widths and Doppler Shifts between Open and Closed Magnetic Structures}",
    journal = {ApJ},
     year = {2024},
     month = may,
    volume = 966,
    pages = {122},    
    doi = {10.3847/1538-4357/ad3424},
     adsurl = {https://ui.adsabs.harvard.edu/abs/2024ApJ...966..122Z/abstract}
}

@ARTICLE{Ryan2021,
   author = {{Ryan}, A. M. and {Gallagher}, P. T. and {Carley}, E. P. and {Brentjens}, M. A. and {Murphy}, P. C. and {Vocks}, C. and {Morosan}, D. E. and {Reid}, H. and {Magdaleni{'c}}, J. and 
   {Brietling}, F. and {Zucca}, P. and {Fallows}, R. and {Mann}, G. and {Kerdraon}, A. and {Halfwerk}, R.},
    title = "{LOFAR imaging of the solar corona during the 2015 March 20 solar eclipse}",
  journal = {A\&A},
     year = {2021},
   volume = 648,
    pages = {A43},
    doi = {10.1051/0004-6361/202039024}, 
   adsurl = {https://ui.adsabs.harvard.edu/abs/2021A%26A...648A..43R/abstract}
}

@ARTICLE{Zhangx2022,
author = {{Zhang}, X. and {Liu}, Y. and {Zhao}, M. and {Liu}, J. and {Elmhamdi}, A. and {Song}, T.},
    title = "{Comparison of the Coronal Green-line Intensities with the EUV Measurements from SDO/AIA}",
    journal = {Res. Astron. Astrophys.},
     year = {2022},
     month = jul,
    volume = 22,
    pages = {075012},
      doi = {10.1088/1674-4527/ac712e},
   adsurl = {https://ui.adsabs.harvard.edu/abs/2022RAA....22g5012Z/abstract}
}

@ARTICLE{Yang2024,
author = {{Yang}, Z. and {Tian}, H. and {Tomczyk}, S. and {Liu}, X. and {Gibson}, S. and {Morton}, R. J. and {Downs}, C.},
    title = "{Observing the evolution of the Sun's global coronal magnetic field over 8 months}",
  journal = {Science},
     year = {2024},
   volume = 386,
    pages = {76},
    doi = {10.1126/science.ado2993}, 
   adsurl = {https://ui.adsabs.harvard.edu/abs/2024Sci...386...76Y/abstract}
}

@ARTICLE{Landi2016,
   author = {{Landi}, E. and {Habbal}, S. R. and {Tomczyk}, S.},
    title = "{Coronal plasma diagnostics from ground-based observations}",
    journal = {J. Geophys. Res.: Space Physics},
      year = {2016},
     month = sep,
    volume = 121,
    pages = {8237},
    doi = {10.1002/2016JA022598},
    adsurl = {https://ui.adsabs.harvard.edu/abs/2016JGRA..121.8237L/abstract}
}

@ARTICLE{Lemen2012,
   author = {{Lemen}, J.~R. and {Title}, A.~M. and {Akin}, D.~J. and {Boerner}, P.~F. and {Chou}, C. and {Drake}, J.~F. and {Duncan}, D.~W. and {Edwards}, C.~G.},
    title = "{The Atmospheric Imaging Assembly (AIA) on the Solar Dynamics Observatory (SDO)}",
  journal = {Sol. Phys.},
     year = 2012,
    month = jan,
   volume = 275,
    pages = {17},
      doi = {10.1007/s11207-011-9776-8},
   adsurl = {http://adsabs.harvard.edu/abs/2012SoPh..275...17L}
}

@ARTICLE{Huangfu2025,
   author = {{Huangfu}, C. and {Fu}, H. and {Li}, B. and {Huang}, Z. H. and {Sun}, M. Z. and {Liu}, W. X. and {Yu}, X. Y. and {Xia}, L. D.},
    title = "{Enhanced Nonthermal Line Broadening inside Coronal Cavities above Solar Prominences Revealed by Spectral Imaging CoronaGraph}",
  journal = {ApJL},
     year = 2025,
    month = dec,
   volume = 995,
    pages = {L71},
      doi = {10.3847/2041-8213/ae2a35},
   adsurl = {https://ui.adsabs.harvard.edu/abs/2025ApJ...995L..71H/abstract}
}

@ARTICLE{Ding2017,
   author = {{Ding}, A. and {Habbal}, S. R.},
    title = "{First Detection of Prominence Material Embedded within a 2 × 10^6 K CME Front Streaming away at 100-1500 km/sec in the Solar Corona}",
    journal = {ApJL},
     year = {2017},
     month = jun,
    volume = 842,
    pages = {L7},    
    doi = {10.3847/2041-8213/aa7460},
     adsurl = {https://ui.adsabs.harvard.edu/abs/2017ApJ...842L...7D/abstract}
}

@ARTICLE{Alissandrakis2021,
  author = {{Alissandrakis}, C. E. and {Gary}, D. E.},
    title = "{Radio Measurements of the Magnetic field in the Solar Chromosphere and the Corona}",
  journal = {Front. Astron. Space Sci.},
     year = {2021},
   volume = 7,
    pages = {77},
    doi = {10.3389/fspas.2020.591075}, 
   adsurl = {https://ui.adsabs.harvard.edu/abs/2021FrASS...7...77A/abstract}
}

@ARTICLE{Ramesh_2010_grip,
       author = {{Ramesh}, R. and {Kathiravan}, C. and {Sastry}, {\relax Ch} V.},
               title = "{Estimation of Magnetic Field in the Solar Coronal Streamers Through Low Frequency Radio Observations}",
      journal = {ApJ},
         year = 2010,
        month = mar,
       volume = {711},
       number = {2},
        pages = {1029-1032},
          doi = {10.1088/0004-637X/711/2/1029},
       adsurl = {https://ui.adsabs.harvard.edu/abs/2010ApJ...711.1029R}
}

@article{Vrsnak_2002,
	author = {{Vrsnak}, B. and {Magdaleni{\'c}}, J. and {Aurass}, H. and {Mann}, G.},
	title = {Band-splitting of coronal and interplanetary type II bursts - II. Coronal magnetic field and Alfv\'en velocity},
	DOI= {10.1051/0004-6361:20021413},
	url= {https://doi.org/10.1051/0004-6361:20021413},
	journal = {A\&A},
	year = 2002,
	volume = {396},
	number = {2},
	pages = {673-682},
}

@ARTICLE{Dulk_&_Mclean_1978,
	author = {{Dulk}, G.~A. and {McLean}, D.~J.},
	title = "{Coronal magnetic fields.}",
	journal = {Sol. Phys.},
	year = 1978,
	month = apr,
	volume = {57},
	number = {2},
	pages = {279-295},
	doi = {10.1007/BF00160102},
	adsurl = {https://ui.adsabs.harvard.edu/abs/1978SoPh...57..279D}
}

@ARTICLE{Sayuf2026b,
       author = {{Sayuf}, Shaik. and {Kathiravan}, C. and {Gireesh}, G. V. S. and {Indrajit V.}, Barve. and {Rajesh}, M. and {Shanmukha Rao}, B. and {Ramesh}, R.},
        title = "{Augmentation of the Gauribidanur radioheliograph for observations of circularly polarized emission from the solar corona at low frequencies}",
      journal = {arXiv e-prints},
          year = 2026,
        month = aug,
          eid = {arXiv:2608.05019},
          doi = {10.48550/arXiv.2608.05019},
archivePrefix = {arXiv},
       primaryClass = {astro-ph.SR}
}

@ARTICLE{Gopalswamy2012,
   author = {{Gopalswamy}, N. and {Nitta}, N. and {Akiyama}, S. and {M{\"a}kel{\"a}}, P. and 
	{Yashiro}, S.},
    title = "{Coronal Magnetic Field Measurement from EUV Images Made by the Solar Dynamics Observatory}",
  journal = {ApJ},
archivePrefix = "arXiv",
   eprint = {1109.2925},
 primaryClass = "astro-ph.SR",
     year = 2012,
    month = jan,
   volume = 744,
      eid = {72},
    pages = {72},
      doi = {10.1088/0004-637X/744/1/72},
   adsurl = {http://adsabs.harvard.edu/abs/2012ApJ...744...72G}
}

@ARTICLE{Bastian2001,
   author = {{Bastian}, T.~S. and {Pick}, M. and {Kerdraon}, A. and {Maia}, D. and 
	{Vourlidas}, A.},
    title = "{The Coronal Mass Ejection of 1998 April 20: Direct Imaging at Radio Wavelengths}",
  journal = {ApJL},
     year = 2001,
    month = sep,
   volume = 558,
    pages = {L65-L69},
      doi = {10.1086/323421},
   adsurl = {http://adsabs.harvard.edu/abs/2001ApJ...558L..65B}
}

@INPROCEEDINGS{Gelfreikh2004,
   author = {{Gelfreikh}, G. B.},
    title = "{Coronal Magnetic Field Measurements Through Bremsstrahlung Emission}",
booktitle = {Solar and Space Weather Radio Physics, ASSL Vol.314},
publisher = {Springer Science + Business Media, Inc.},
   editor = {{Gary}, D. E. and {Keller}, C. U.},
   year = 2004,   
    month = sep,
    pages = {314},
   doi = {10.1007/1-4020-2814-8_6},
   adsurl = {http://adsabs.harvard.edu/abs/2004ASSL..314..115G}
}

@INPROCEEDINGS{White2004,
   author = {{White}, S. M.},
    title = "{Coronal Magnetic Field Measurements Through Gyroresonance Emission}",
booktitle = {Solar and Space Weather Radio Physics, ASSL Vol.314},
publisher = {Springer Science + Business Media, Inc.},
   editor = {{Gary}, D. E. and {Keller}, C. U.},
   year = 2004,   
    month = sep,
    pages = {89},
      doi = {10.1007/1-4020-2814-8_5},
   adsurl = {http://adsabs.harvard.edu/abs/2004ASSL..314...89W}
}

@INPROCEEDINGS{Ryabov2004,
   author = {{Ryabov}, B.},
    title = "{Coronal Magnetic Field Measurements Through Quasi-Transverse Propagation}",
booktitle = {Solar and Space Weather Radio Physics, ASSL Vol.314},
publisher = {Springer Science + Business Media, Inc.},
   editor = {{Gary}, D. E. and {Keller}, C. U.},
   year = 2004,   
    month = sep,
    pages = {135},
      doi = {10.1007/1-4020-2814-8_7},
   adsurl = {http://adsabs.harvard.edu/abs/2004ASSL..314..135R}
}

@ARTICLE{Kakinuma1962,
   author = {{Kakinuma}, T. and {Swarup}, G.},
    title = "{A Model for the Sources of the Slowly Varying Component of Microwave Solar Radiation}",
  journal = {ApJ},
     year = 1962,
    month = nov,
   volume = 136,
    pages = {975},
      doi = {10.1086/147450},
   adsurl = {https://ui.adsabs.harvard.edu/abs/1962ApJ...136..975K/abstract}
}

@article{Sastry_2009,
	doi = {10.1088/0004-637x/697/2/1934},
	url = {https://doi.org/10.1088/0004-637x/697/2/1934},
	year = 2009,
	month = {may},
	publisher = {American Astronomical Society},
	volume = {697},
	number = {2},
	pages = {1934--1939},
	author = {{Sastry}, {\relax Ch} V.},
	title = {{POLARIZATION} {OF} {THE} {THERMAL} {RADIO} {EMISSION} {FROM} {OUTER} {SOLAR} {CORONA}},
	journal = {ApJ},
}

@article{Agarwal2026,
	author = {{Agarwal}, A. and {Mishra}, W. and {Ramesh}, R. and {Muthu Priyal}, V. and {Zhukov}, V. N. and {Singh}, J. and {Gopalswamy}, N.},
	title = {Coordinated Coronagraphic Observations from Proba-3 and Aditya-L1: Investigating CME Energetics in the Inner Corona},
	journal = {A\&A (in press)},
	year = 2026,
	}
\end{document}